\documentclass[ aip,
amsmath,amssymb,
reprint %
]{revtex4-1}

\usepackage{graphicx,color}
\usepackage{dcolumn}
\usepackage{bm}
\usepackage[section]{placeins}
\usepackage{color}

\begin{document}

\title{Taylor line swimming in microchannels and cubic lattices of obstacles}


\author{Jan L. M\"unch}
 \affiliation{Institut f\"ur Theoretische Physik, Technische Universit\"at Berlin, Hardenbergstr. 36, D-10623 Berlin, Germany}
 \author{Davod Alizadehrad}%
\affiliation{Institut f\"ur Theoretische Physik, Technische Universit\"at Berlin, Hardenbergstr. 36, D-10623 Berlin, Germany}
\affiliation{Forschungszentrum J\"ulich, Wilhelm-Johnen-Stra{\ss}e, D-52425 J\"ulich, Germany}
\author{Sujin Babu}
\affiliation{Department of Physics, Indian Institute of Technology Delhi, Hauz Khas, New Delhi-110016, India}
\author{Holger Stark}
\email{Holger.Stark@tu-berlin.de}
\homepage{http://www.itp.tu-berlin.de/stark}
\affiliation{Institut f\"ur Theoretische Physik, Technische Universit\"at Berlin, Hardenbergstr. 36, D-10623 Berlin, Germany}

\date{\today}

\begin{abstract}
Microorganisms naturally move in microstructured  fluids. Using the 
simulation method of multi-particle collision dynamics,
we study an undulatory Taylor line swimming in a two-dimensional microchannel and in a cubic lattice of obstacles, 
which represent simple forms of a microstructured environment. In the microchannel the Taylor line swims at an acute angle 
along a channel wall with a clearly enhanced swimming speed due to hydrodynamic interactions with the bounding wall.
While in a dilute obstacle lattice swimming speed is also enhanced, a dense obstacle lattice gives rise to \emph{geometric swimming}.
This new type of swimming is characterized by a drastically increased swimming speed. 
Since the Taylor line has to fit into the free space of the obstacle lattice, the swimming speed is close to the phase velocity 
of the bending wave traveling along the Taylor line.
While adjusting its swimming motion within the lattice, the Taylor line chooses a specific
swimming direction, which we classify by a lattice vector. 
When plotting 
the swimming velocity versus the magnitude of the lattice vector, all our data collapse on a single master curve. 
Finally, we also report more complex trajectories
within the obstacle lattice.

\end{abstract}

\pacs{Valid PACS appear here}

\maketitle

\section{Introduction}

The motility of microorganisms in their liquid environment is important in various biological 
processes \cite{Lauga2009review}.
Microorganisms move in the low-Reynolds-number regime, where viscous forces dominate 
over inertia \cite{empurcel1976}.
They have developed various swimming strategies to cope with the strong viscous forces \cite{empurcel1976} including beating flagellar appendages of sperm cells \cite{Lighthill1976,Suarez2006},
metachronal waves of collectively moving cilia on the cell surface of a paramecium \cite{Gibbons1981},
rotating helical flagella in \emph{E.coli} \cite{berg1991,Berg2008,Vogel2013,Tapan2015,Hu2015},
%
%
%
%
%
and periodic deformations of the whole cell body \cite{Heddergott2012,Berman2013,Bilbao2013}.
A first expression for the swimming speed of a simplified 
flagellar model was given by Taylor  in 1951 \cite{taylor1951}. 
In this model a prescribed bending wave moves along a filament, which we call \emph{Taylor line} in the following.
A recent study with the Taylor line showed hydrodynamic
phase locking of multiple flagellas \cite{Elfring2009Lauga} 
and Ref. \cite{MontegroJohnson2014Lauga} determined
the optimal shape of a large amplitude wave.
These insights into biological swimming mechanisms in Newtonian liquids 
inspired
studies of artificial swimmers in unbounded \cite{dreyfus2005,Gauger2006} 
as well as bounded \cite{Rojman2009,Crowdy2011} fluids.

Following the seminal experiments of Rothschild in 1963 \cite{Rothschild1963},
artificial microchannels 
have extensively been
used to investigate the influence
of bounding walls on locomotion 
\cite{Berke2008,Murray2009,Elgeti09,Guanglai2009,Guanglai2011,Kiori2012,Zoettl12,Zoettl13,Elgetti13,
GaoJin2014,SpangnolieLauga2012,Uppaluri2012,Bilbao2013,Rusconi14,Lauga2015}.
Hydrodynamic interactions of sperm cells with channel walls 
\cite{lopez2008,Elgeti2010,Denissenko2012,kantsler2014surface,Zoettl15,nosrati2015}
and with other cells \cite{Yang2008} are of special interest in reproductive medicine. 

In vivo the motility of protozoa and small eukaryotic organisms is influenced by 
obstacles in the liquid environment such as 
cells \cite{Mota2001,Vanderberg2004,Engstler2007}
and proteins \cite{Schneider1974,berg1979,Shen2011,Liu2011, Martinez2014},
but also studies with artificially produced posts exist
\cite{Park2008,Majmudar2012,Heddergott2012}.
Not only the shape of the obstacles
is important but  they also can make the liquid environment viscoelastic.
Examples in nature of biological or medical 
relevance include microorganisms in 
soil \cite{Park2008, Majmudar2012}, 
in blood \cite{Mota2001,Vanderberg2004,Engstler2007}, or in
mucus \cite{Voynow2009,Johanson2010, Hansson2012, Druart2012}. 
The mucus of the cervix uteri, for example, consists of a dense polymer network. 
This polymer network induces a hydrodynamic sorting process.
Sperms cells with 
normal swimming motion are able to
pass the network whereas 
for defective sperm cells the mucus is 
hardly penetrable
\cite{Suarez2006}.
Model swimmers with large-amplitude deformations of their 
driving filament show speed enhancement
in viscoelastic fluids 
\cite{Theran2010Shelley,Riley14},
while for small-amplitude deformations viscoelasticity 
hinders faster
swimming \cite{Lauga2007,Fu2007,Fu2009,Theran2010Shelley,Liu2011}.
Experiments with \textit{C. elegans}
in viscoelastic fluids confirm the 
prediction
of slower swimming
\cite{Juarez2010,Shen2011}.

In 1979 L. Turner and H. C. Berg suggested that the geometric 
constraints of polymer networks in viscoelastic fluids can drastically enhance
the swimming speed of microorganisms \cite{berg1979}.
Based on experimental observations with helical bacteria they formulated the following picture.
When
rotating about their helical axis, bacteria with helical shape move through a polymeric liquid like through a quasi-rigid 
medium and similar to a corkscrew driven into cork. So, in the ideal case, after each full rotation the bacterium would proceed
by one full pitch length. In this paper we will investigate another type of this \emph{geometrical swimming} by studying
the Taylor line in a cubic lattice of obstacles.

A typical example for obstacles in nature are erythrocytes or red blood cells.
The African trypanosome, the causative agent of the sleeping sickness, swims faster
in the crowded environment of blood and thereby removes surface-bound antibodies
with the help of hydrodynamic drag forces \cite{Engstler2007}.
In this way, the parasite evades the immune response of its host. The motility of the African trypanosome
in a Newtonian liquid was investigated in bulk fluid by computer modeling
\cite{SujinSchmeltzer2012,Babu2012,Alizadehrad2015} and in Poiseuille flow \cite{Uppaluri2012}.
Blood is a complex viscoelastic liquid containing a large amount of cellular components,
which gives blood a non-Newtonian character.
Its viscosity depends on the volume fraction of erythrocytes 
(hematocrit), shear rate, and temperature \cite{Wells1962,Peter1964}.
In order to understand the geometrical constraints of erythrocytes for the motility of the trypansome or 
how other obstacles influence the swimming of sporozoites or \textit{C. elegans}, 
more controlled experiments
were conducted. They use either
suspended colloids 
\cite{Jung2010,Juarez2010} 
or fabricated lattices of posts
\cite{Heddergott2012,Battista2014, Park2008, Majmudar2012, Johari2013}. 

In lab-on-chip devices obstacle lattices are used to separate trypanosomes from 
erythrocytes with the idea to diagnose the sleeping sickness in an early stage \cite{Holm2011}.
Trypanosomes swimming in these lattices
show a motility much more comparable to their
in vivo motility due to interactions with the obstacles
\cite{Heddergott2012}. Similarly,
Park \emph{et al.} found 
that \textit{C. elegans} swims up to ten times faster in an obstacle lattice 
compared to its swimming speed
in bulk fluid \cite{Park2008}.
The speed-up depended
on the lattice spacing.
A combined experimental and numerical study by Maj\-mudar \textit{et al.} 
with an undulatory swimmer such as \textit{C. elegans}
showed that most of the characteristics of this
new type of swimming in an array of
micro pillars can be explained by a
mechanical model for the swimmer \cite{Majmudar2012}.
It does not need any
biological sensing or behavior. 

In this paper we present a detailed hydrodynamic study of an undulatory Taylor line swimming in a two-dimensional
microchannel and in a cubic lattice of obstacles. We use the method of multi-particle collision dynamics 
for simulating the hydrodynamic flow fields \cite{Gompper2008rev}.
In the microchannel the Taylor line swims at an acute angle along a channel wall with a clearly enhanced
swimming speed.
In a dilute obstacle lattice swimming speed is also enhanced due to hydrodynamic interactions with the 
obstacles similar to a study by Leshansky \cite{Leshansky2009}.
Moving the obstacles closer together (dense obstacle lattice), the undulatory Taylor line has to fit into
the free space of the obstacle lattice, where it performs geometric swimming. Here, the swimming speed is close to the wave velocity
of the bending wave traveling along the Taylor line. In this regime, we classify the possible swimming directions by lattice 
vectors. When plotting the ratio of swimming and wave velocity versus the magnitude of the lattice vector (effective lattice constant), 
all our data collapse on a single master curve. This demonstrates the regime of geometric swimming. We also illustrate more 
complex trajectories.

The article is structured as follows.
In Sec. \ref{chap_methodes} we introduce our computational methods including the method of multi-particle
collision dynamics and the implementation of the Taylor line.
In Sec.\ \ref{sec_results} we calibrate the parameters of the Taylor-line model by studying its swimming 
motion in the bulk fluid. 
In Secs.\ \ref{sec.micro} and \ref{sec.micro} we review the respective results for swimming in the microchannel and
in the obstacle lattice. Sec.\ \ref{sub_discus} closes with a summary and conclusions.

\section{Computational methods}    \label{chap_methodes}
\subsection{Multi-particle collision dynamics}
\label{sub_mpcd}

We employ the method of multi-particle collision dynamics (MPCD) to simulate the Taylor line in its fluid 
environment \cite{malevanets2000,Kapral1999}. MPCD uses point particles of mass $m_0$ as coarse-grained fluid
particles. Their dynamics consists of a ballistic streaming and a collision step, which locally conserves momentum.
Therefore, the resulting flow field satisfies the Navier-Stokes equations but also inherently includes thermal fluctuations
\cite{Gompper2008rev}.

In the streaming step the positions $\vec{r}_i$ of all fluid particles are updated according to
\begin{equation}
 \vec{r}_i(t+\Delta t_c) = \vec{r}_i(t)+\vec{v}_i(t)\Delta t_{c} \, ,
\end{equation}
where  $\vec{v}_i$ is the particle velocity and $\Delta t_c$ the MPCD time step between collisions\ \cite{Bolintineanu2012}.

After each streaming step the fluid particles are sorted into quadratic collision cells of linear dimension $a_0$, so that on
average each cell contains $N$ particles with mass $M = Nm_0$. In each cell we redistribute the particles' velocities
following a collision rule, for which we choose the Anderson thermostat with additional angular momentum conservation 
\cite{Bolintineanu2012}. At first we calculate the total momentum, $\vec{P}_{cell} = m_0 \sum_{i\in cell}\vec{v}_i$, 
of each collision cell. Then, we assign to each velocity component of a particle relative to the mean velocity 
$\vec{P}_{cell} / M$ a random component $v_{i,rand}$ from a Gaussian distribution with variance $k_B T/m_0$. 
Here, $T$ is temperature and $k_B$ the Boltzman constant. Using the mean random momentum
$\vec{P}_{rand} = m_0 \sum_{i\in cell} \vec{v}_{i,rand}$ of each cell, we determine the new particle velocities after
the collision:
\begin{align}
\vec{v}_{i,new}^{C}=\frac{\vec{P}_{cell}}{M} + \vec{v}_{i,rand} - \frac{ \vec{P}_{rand}(t)}{M} \, .
\label{AT}
\end{align}
This collision rule conserves linear momentum but not angular momentum \cite{Gompper2008rev}.
To keep the latter constant, we note that during the collision step the fluid particles have fixed distances. 
Therefore, one can apply a rigid body rotation, $ \Delta \vec{\omega} \times \vec{r}_i $, to replace the
new velocities $\vec{v}_{i,new}^{C}$ by
\begin{align}
 \vec{v}_{i,new} = \vec{v}_{i,new}^{C} -\Delta \vec{\omega} \times \vec{r}_i\text{.}
\end{align}
Here, the angular velocity is
\begin{align}
\Delta \vec{\omega}  = 
m_0 \Theta^{-1} 
\sum_{i \in \mathrm{cell}}  \vec{r}_{i} \times (\vec{v}_{i,rand}-\vec{v}_i) \, ,
\label{omega}
\end{align}
where 
$\Theta=  m_0 \sum_{i \in cell} |\vec{r}_i|^2 $ is the
moment of inertia of the particles in the cell.
This rule restores angular momentum conservation keeping linear momentum constant. By definition, the collision rule 
based on the Anderson thermostat also keeps the temperature constant. To restore Galilean invariance and the 
molecular chaos assumption, we always apply a random grid shift when defining the collision cells and take the 
shift from the interval $[0, a_0]$ \cite{Ihle2001,Ihle2003}.
Transport coefficients of the MPCD fluid can be found in Ref. \cite{Noguchi2008}.

In the following, we will measure quantities in typicial MPCD units. We will use the linear dimension
of the collision cell $a_0$ as a unit for lengths, energies are measured in units of $k_BT$, and 
mass in units of $m_0$. Then the time unit becomes $\tau_0 = a_0 \sqrt{m_0/k_BT}$ \cite{Padding2006}.
In this unit, our time step between collisions is always chosen as $\Delta t_c = 0.01$.

\subsection{No-slip boundary condition: Bounce-back rule and vir\-tual particles}

\begin{figure}
 \includegraphics[width=0.49\textwidth]{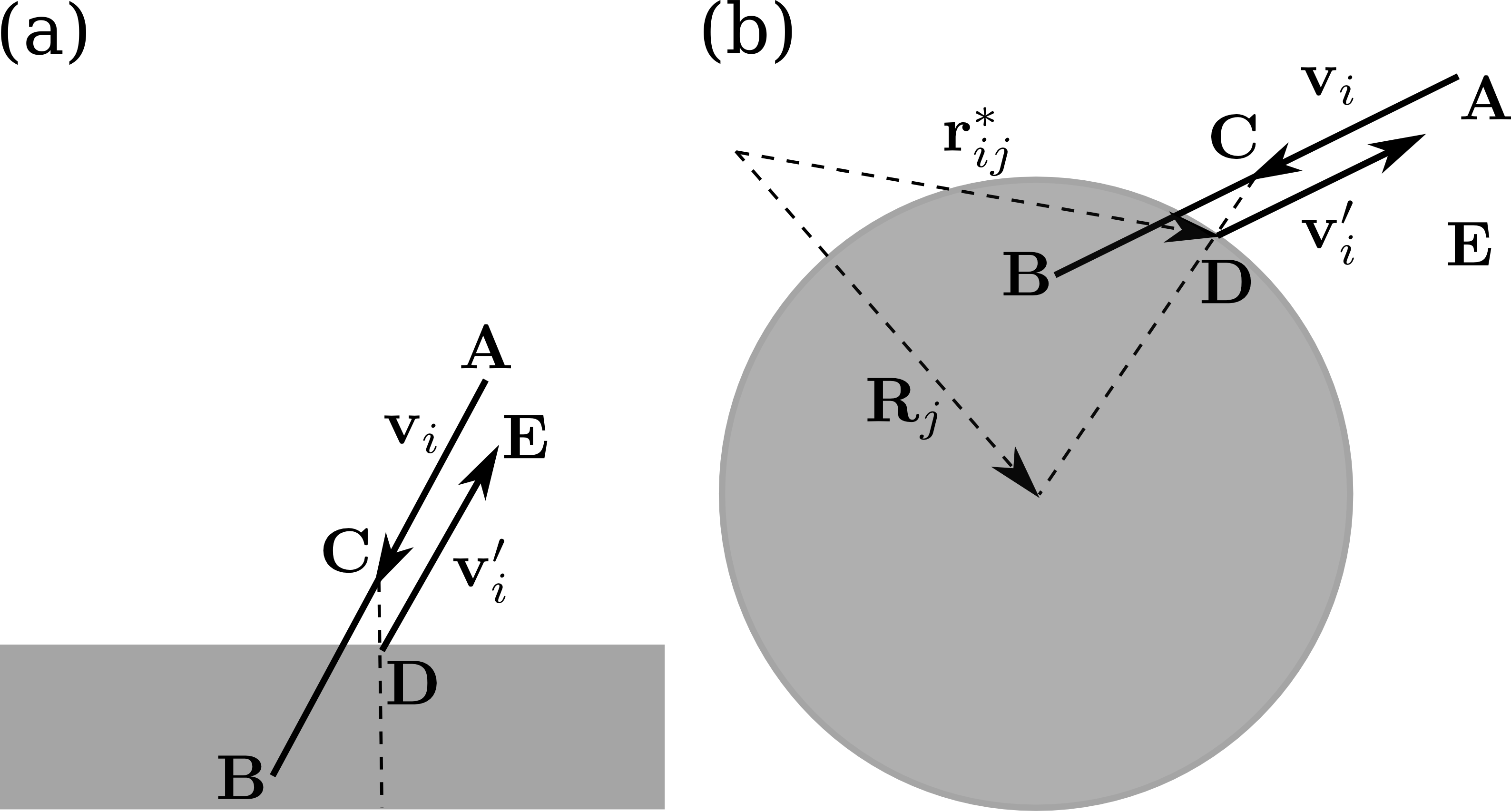}
\caption{
Sketch of the bounce-back rule at (a) a channel wall and (b) an obstacle.
Particle positions during implementation of the rule are denoted by capital letters
and explained in the main text. The velocities before and after the bounce are
denoted by $\vec{v}_i$ and $\vec{v}^{\,\prime}_i = -\vec{v}_i$, respectively.
}
\label{fig_bounce_rules}
\end{figure}

At bounding walls fluid flow obeys the no-slip boundary condition. To implement it within the MPCD method,
we let the effective fluid particles interact with channel walls or obstacles using the bounce-back rule, 
see Fig. \ref{fig_bounce_rules}.
When a fluid particle moves into
an obstacle or a channel wall during the streaming step (position B), we invert 
the velocity $\vec{v}^{\,\prime}_{i,} = -\vec{v}_{i}$ and let the particle stream to position C during half the collision time:
\begin{align}
 \vec{r}_i(t+\Delta t_c/2) = \vec{r}_i(t)+\vec{v}^{\,\prime}_{i}(t)\Delta t_{c}/2 \, .                                                                                                                                                                                                                                                       \end{align}
Then, we move this particle to the closest spot on the obstacle surface or channel wall (position D) and
let it stream with the reversed velocity during half the collision time to position E.

\begin{figure}
\includegraphics[width=0.45\textwidth]{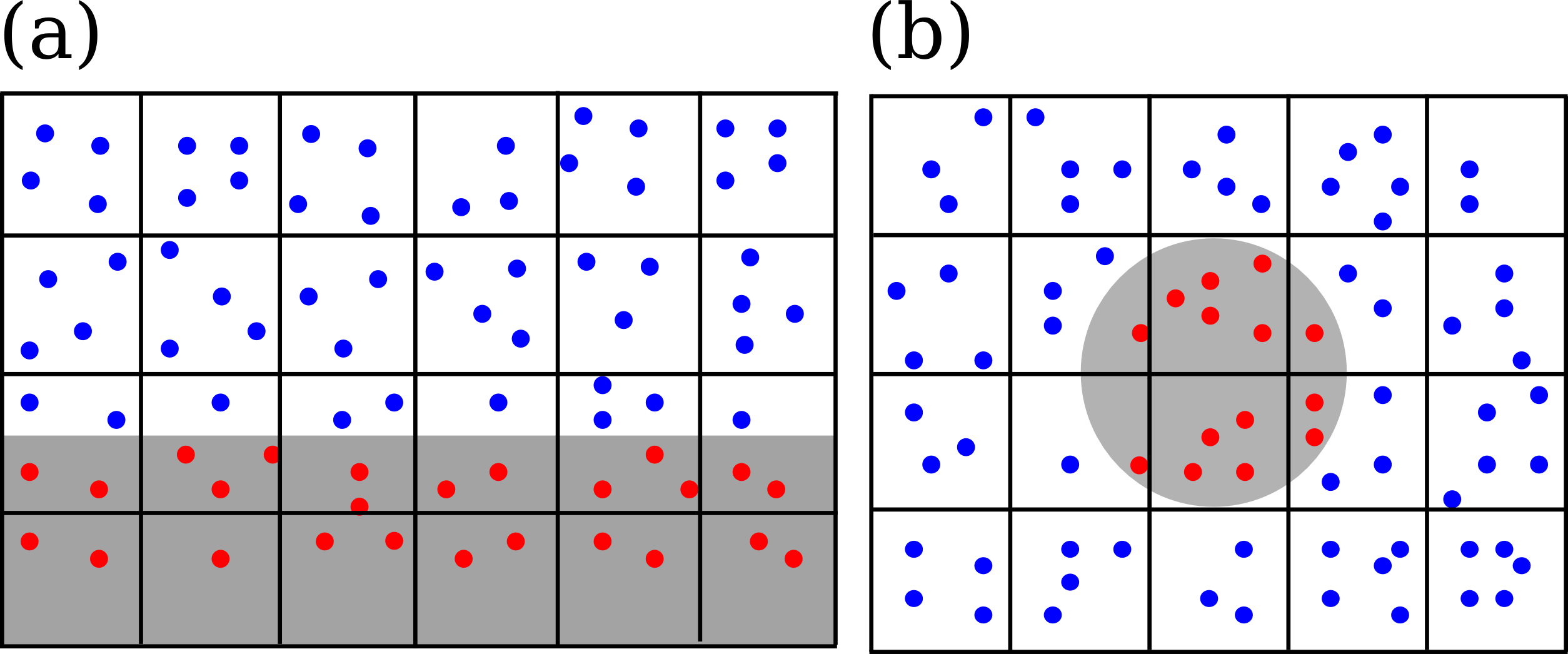}
\caption{
Coarse-grained fluid particles (blue) and virtual particles (red) close to (a) a channel wall and (b) an obstacle,
which are represented by gray areas. Both figures show the lattice of collision cells. The fluid particles cannot
penetrate into the gray areas.
}
\label{fig_ghost}
\end{figure}

In addition, the no-slip boundary condition is improved using virtual particles inside a channel wall or
an obstacle, see Fig. \ref{fig_ghost}. We uniformly distribute virtual particles (red dots in Fig. \ref{fig_ghost}) in the
areas of the collision cells, which extend into the channel wall or obstacles. The velocity components are
chosen from a Gaussian distribution with variance $k_BT/m_0$.
The virtual particles also take part in the collision step. So, close to bounding walls one has the same average
number of particles in a collision cell as in the bulk. Both rules together implement the no-slip boundary condition
at a bounding surface in good approximation\ \cite{Lamura2002, Bolintineanu2012}.

\subsection{A discrete model of the Taylor line}   \label{chap_the_model}

\begin{figure}
\includegraphics[width=0.5\textwidth]{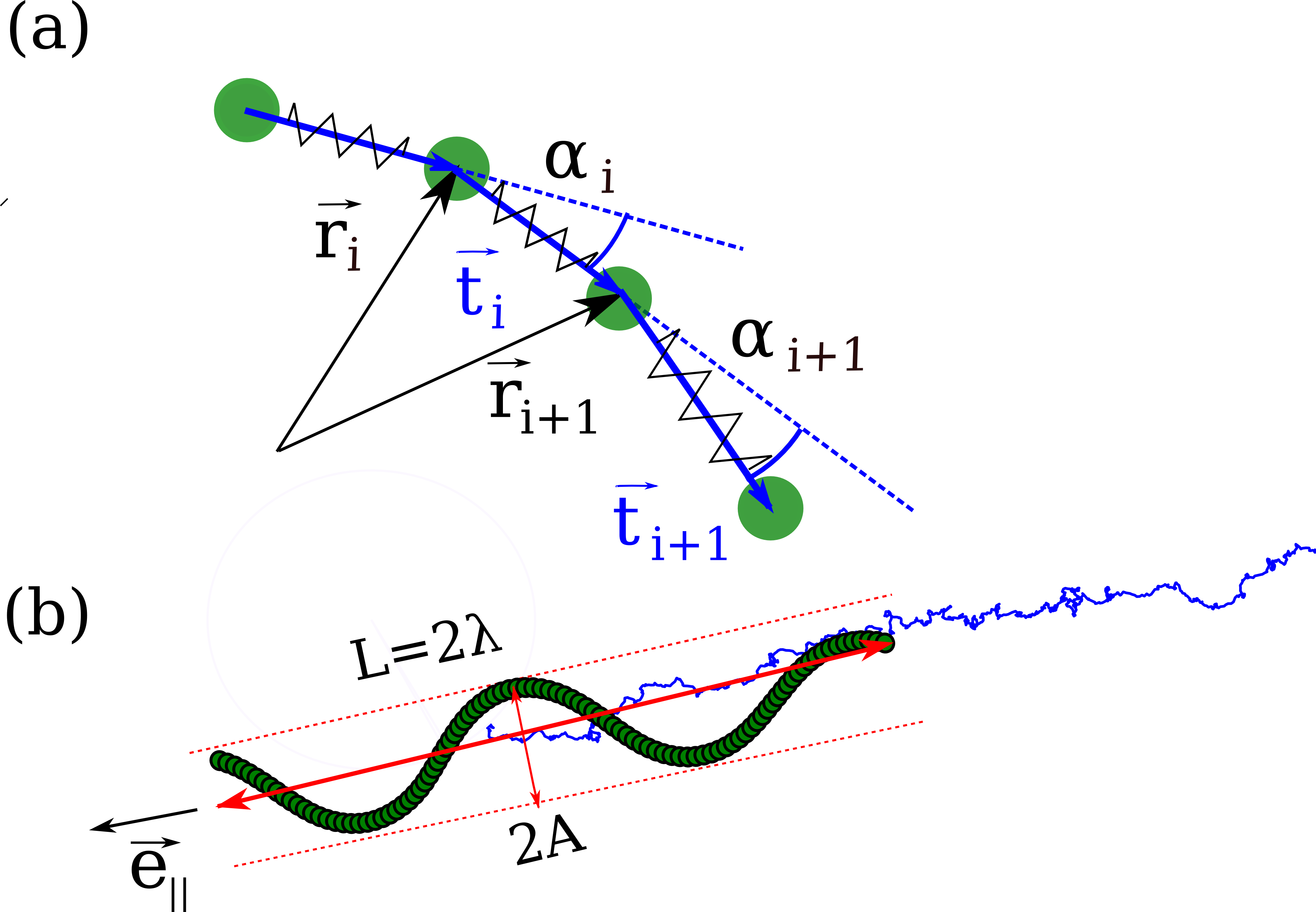}
\caption{(a) The Taylor line is modeled as a bead-spring chain, where $ \vec{r}_i$ gives the bead position.
The tangential vector $\vec{t}_i = \vec{r}_{i+1} - \vec{r}_i$ connects two neighboring beads and is not normalized
to one. The angles $\alpha_i$ between the tangential vectors are used to define the sinusoidal bending wave running
along the Taylor line.
(b) Snapshot of the Taylor line, which swims along the unit vector $\vec{e}_{\|}$ in a bulk fluid with superimposed
thermal diffusion. The blue line represents the center-of-mass trajectory. 
The end-to-end distance of the Taylor line or its length along $\vec{e}_{\|}$ is $L= 2\lambda$,
where $\lambda$ is the the wavelength of the bending wave along $\vec{e}_{\|}$
and $A$ its amplitude.
}
\label{beads}
\end{figure}

The Taylor line propels itself by running a sinusoidal bending wave along its contour line.
Figure\ \ref{beads}(a) shows how we discretize the Taylor line by a bead-spring chain with $N$
beads each of mass $m = 10\,m_0$. The beads at positions $\vec{r}_i$ interact with each other by a spring 
and a bending potential. The spring potential implements Hooke's law between nearest 
neighbors \cite{Gauger2006}, 
\begin{align}
    V_H = \frac{D}{2} \sum_{i=1}^{N-1} ( | \vec{t}_{i} | - l_0 )^2 \, .
\end{align}
Here $l_0 = 1/2\,a_0$ is the equilibrium distance between the beads and  $| \vec{t}_i | = | \vec{r}_{i+1}-\vec{r}_i |$
the actual distance, where $ \vec{t}_i$ denotes the tangent vectors.
The contour length of the bead-spring chain,
\begin{align}
L_c = \sum_{i=1}^{N-1} |\vec{t}_i|\approx (N-1)l_0 = (N-1)a_0 / 2 \, ,
 \label{eq_arclegth}
\end{align}
is approximately constant. We choose a large spring constant $D = 10^6$ to ensure that
deviations from the equilibrium distance $l_0$ between the beads are smaller than
$0.002 l_0$.
Finally, the spring force acting on bead $i$ is
\begin{align}
   \vec{F}_i^{H} = - \vec{\nabla}_i V_H = - D (l_i - l_0) \vec{t}_i + D (l_{i+1} - l_0) \vec{t}_{i+1} \, .
\end{align}

The bending potential creates a sinusoidal bending wave that runs along the Taylor line. It was also used in two-dimensional
studies of swimming sperm cells \cite{Yang2008}
 and in simulations of the African typansome \cite{SujinSchmeltzer2012,Babu2012}.
The bending potential has the form:
\begin{align}
 V_{B} =  \frac{\kappa}{2} \sum_{i=1}^{N-1}[\vec{t}_{i+1} - R(\alpha_i)\vec{t}_i]^2 \, ,
\end{align}
where $\kappa = p k_bT$ is the bending rigidity and $p$ the persistence length \cite{nelson2008}. 
The rotation matrix $R(\alpha)$ rotates the tangential vector by an angle $\alpha$ about the normal of the plane,
so the equilibrium shape of the Taylor line is not straight but bent. For the rotation angle at bead $n$ we choose
$\alpha_n =l_0 c(n,t)$, where the equilibrium curvature, 
\begin{align}
 c(n,t) = b\, \text{sin}[\phi(t,n)] = b\, \text{sin}[2\pi (\nu t + n l_0 / \lambda_c)] ,
\label{eq.bending}
\end{align}
is a function of the position of bead $n$ on the Taylor line ($n \in \{1,N\}$) and time $t$.
It creates the sinusoidal bending wave running along the Taylor line with wavelength $\lambda_c$ 
(measured along the contour)
and an amplitude $A$ controlled by the
parameter $b$. 
Unless stated otherwise, 
we choose the ratio of persistence to contour length as $p/L_c = 5 \cdot 10^3$
to ensure that bending forces are much stronger than thermal forces,
in order to induce directed swimming \cite{Yang2008}. 
This is investigated in more detail in Sec.\ \ref{sec_results}.

From the bending potential we derive a bending force acting on bead $j$:
 \begin{eqnarray}
   \vec{F}_j^{B} & = & - \vec{\nabla}_j V_B =  \kappa \big(  [\vec{t}_{j-1} - R(\alpha_{j-2}) \vec{t}_{j-2}]\nonumber\\
& &    +[\vec{t}_{j}-\vec{t}_{j-1}+R^T(\alpha_{j-1})\vec{t}_j-R(\alpha_{j-1})\vec{t}_{j-1}]\\
& &    +[\vec{t}_{j} - R^T(\alpha_j)\vec{t}_{j+1}]\big) \, \nonumber
\end{eqnarray}
where $R^T(\alpha_j)$ means transposed matrix.
Then, the total force $ \vec{F}_i = \vec{F}_i^{H} + \vec{F}_i^{B} $ determines the dynamics of the Taylor line. In our
simulations we update the positions of the beads during the streaming step using the velocity Verlet algorithm with 
time step $\delta t = 0.01\Delta t_{c}$\ \cite{Babu2012}. In addition, the beads with mass $m = 10m_0$
participate in the collision step and the components of their random velocities $\vec{v}_{i,rand}$ are chosen 
from a Gaussian distribution with $k_BT /10m_0$. The beads 
thereby interact with the fluid particles which ultimately couples the Taylor line to the fluid environment.
Note, since the beads of the Taylor  line have a different mass than the fluid particles, in all the formulas
of Sec.\ \ref{sub_mpcd} one has to replace $m_0 \sum_{i \in \mathrm{cell}}  \ldots$ by $\sum_{i \in \mathrm{cell}}  m_i \ldots $,
where $m_i$ is the mass of either the fluid particles or the Taylor line beads. The latter
also interact with channel walls or obstacles by the bounce-forward rule, which is very similar to the bounce-back rule used for the fluid particles.
Upon streaming into an obstacle or wall, we place the particle onto position D, 
see Fig.~\ref{fig_bounce_rules}.
However, in contrast to the bounce-back rule, only the velocity component of the bead orthogonal to the surface is inverted.
This ensures that the Taylor line can slip along a surface.

We introduce the normalized end-to-end vector of the Taylor line,
\begin{align}
 \vec{e}_{||} = \frac{1}{|\sum_{i =1}^{N-1}\vec{t}_i|}\sum_{i=1}^{N-1}\vec{t}_i  \, ,
 \label{eq_j}
\end{align}
to quantify the mean swimming direction and denote the end-to-end distance by $L$.
Unless mentioned otherwise, we always fit two complete bending wave trains onto the Taylor line,
meaning $L = 2\lambda$, where $\lambda$ is the wavelength measured along $ \vec{e}_{||}$
[see Fig. \ref{beads} (b)]. Note that $\lambda$ is different from the wavelength $\lambda_c$ along the contour
introduced in Eq.\ (\ref{eq.bending}). In the following, we will vary the amplitude $A$ of the bending wave keeping
the end-to-end distance with $L = 2\lambda$ fixed. Therefore, we always have to adjust the contour length of
the Taylor line by adding or removing some beads. Typically, we use Taylor lines with $L=42 a_0$ and 
the number of beads ranges from $N= 88$ to $125$.

\section{Taylor line in the bulk fluid}   \label{sec_results}

\begin{figure}
\includegraphics[width=0.49\textwidth]{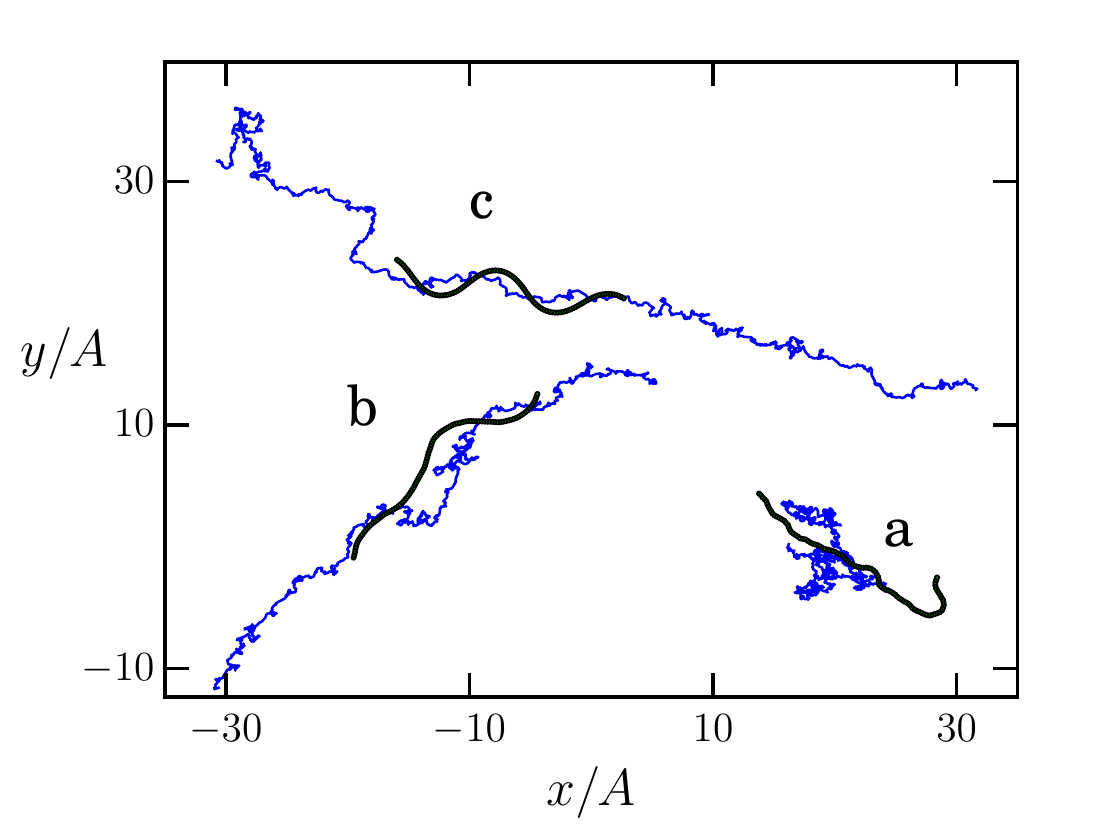}
\caption{Taylor line (chain of green dots) swimming and diffusing in a bulk fluid at different
persistence lengths normalized by the chain length: (a) $p/L_c = 1$, (b) $p/L_c=10$, and (c) $p/L_c=500$.
The blue curve represents the center-of-mass trajectory and the chain of green dots shows a typical snapshot.
The different trajectories are discussed in the main text.
}
\label{fig_stiff_trajec_1} 
\end{figure}

In the following we 
discuss the swimming velocity of the Taylor line as a function of the dimensionless persistence 
length $p / L_c$. 
Thermal fluctuations noticeably bend an elastic line on lengths comparable to the persistence length.
So, in our case the Taylor line should have the form of a sine wave when $p$ is much larger than its contour length $L_c$.
In addition, the Taylor line performs translational and rotational Brownian motion as thermal fluctuations are inherently
present in the MPCD fluid.
All this is visible in Fig.~\ref{fig_stiff_trajec_1}. In case (a) with $p / L _c= 1$ the Taylor line is too sloppy and the
bending wave cannot develop. Only thermal motion of the center of mass occurs (blue line), reminiscent of a Brownian particle.
In case (b) with $p / L_c = 10$ the bending wave is clearly visible, although still distorted by thermal fluctuations, and
the Taylor line 
exhibits persistent motion. The Taylor line has a fully undistorted, sinusoidal contour in case (c) 
at $p / L_c = 500$. The trajectory of the center of mass shows directed swimming superimposed by Brownian motion. 
The total displacement over a complete simulation run 
is larger compared to (b) and the Taylor line has reached its maximum propulsion speed.

\begin{figure} 
 \includegraphics[width=0.49\textwidth]{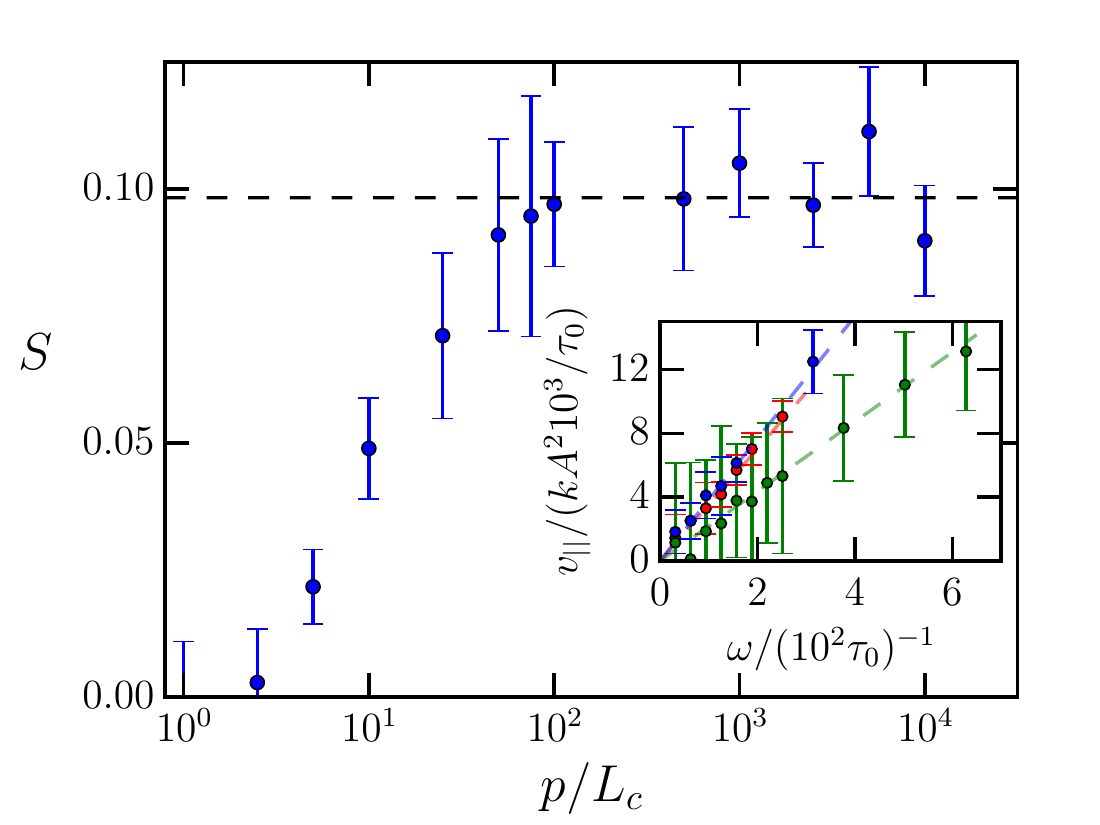}
\caption{
Stroke efficiency $S$ versus dimensionless persistence length $p/L_c$
of the Taylor line. 
The wave frequency is $\nu = 0.003 / \tau_0$ and the amplitude to wavelength ratio is $ A / \lambda = 0.14$.
The error bar shows the standard deviation of a time average over a simulation period of $3000 / \tau_0$.
The dashed line is a linear fit of the last 8 data points.
The inset shows the swimming velocity $\langle v_{\|} \rangle$ in units of $ kA^2 / \tau_0$ as a function of $\omega \tau_0$ 
for different values of $A/\lambda$. Green: $A/\lambda = 0.04$, blue: $A/\lambda = 0.1$, red: $A/\lambda = 0.14$.
The dashed lines are linear fits.
}
\label{fig_speed_bending_stiff} 
\end{figure}
 
To discuss directed swimming more quantitatively, we introduce the swimming velocity 
$v_{||} = d\vec{r}\cdot \vec{e}_{||}/\Delta t$, where we project the center-of-mass displacement $d\vec{r}$ during time 
$\Delta t$ onto the mean direction of the Taylor line defined in Eq.\ (\ref{eq_j}) and indicated in Fig.\ \ref{beads}(b).
We then define the stroke efficiency
\begin{equation}
 S = \frac{\langle v_{||}\rangle}{c} = \frac{\langle v_{||}\rangle}{\lambda \nu} \, .
\label{eq_stroke}
\end{equation}
It compares the mean swimming speed, averaged over the whole swimming trajectory, with the phase velocity $c$,
at which the bending wave travels along the Taylor line. Then, $S=1$ indicates optimal swimming of the Taylor line.
In three dimensions this situation is similar to a corkscrew screwed into the cork. It moves at a speed that equals the
phase velocity of the helical wave traveling along the rotating corkscrew.

In Fig.\ \ref{fig_speed_bending_stiff} we plot the stroke efficiency $S$ versus persistence length $p/L_c$.
For $p=L_c$ the stroke efficiency is approximately zero as already observed from the trajectory (a) in Fig.\ \ref{fig_stiff_trajec_1}.
The efficiency $S$ increases nearly linearly in $\log(p/L_c)$ until at ca. $p/L_c = 10^2$ it reaches a plateau value.
A linear fit gives the plateau value $S_0=0.098$ typical for low Reynolds number swimmers.
For example, for \emph{C. Elegans} studied in Ref.\ \cite{Park2008} we estimate $S = 0.12$.
In the following we always use the persistence length $p/L_c = 5 \cdot 10^3$ to be on the safe side. 
 
Within resistive force theory, one derives for the swimming speed of the Taylor line in the limit of $A \ll \lambda$:
\begin{align}
\langle v_{||}\rangle = \frac{\xi_{\perp}-\xi_{||}}{2\xi_{||}}\omega k A^2 \, ,
 \label{eq_ana_speed}
\end{align} 
with the wave number $k=2\pi / \lambda$ and angular frequency $\omega = 2\pi \nu$. The parameters $\xi_{\perp}$ and
$\xi_{\|}$ are the respective local friction coefficients per unit length for motion perpendicular and parallel to the local tangent
\cite{Lauga2009review}.
Originally, Taylor used $\xi_{\perp} = 2 \xi_{\|}$ valid for an infinitely long filament.
We are able to reproduce the linear relationship between swimming speed 
$\langle v_{||} \rangle$ and $\omega$ in our simulations (see inset of Fig.\ \ref{fig_speed_bending_stiff}).
Whereas $A/\lambda = 0.1$ (blue) and $0.14$ (red) confirm the expected scaling with $kA^2$, the straight line for
$A = 0.04 \lambda = 0.9 a_0$ deviates from it, possibly because the amplitude is too small to be correctly
resolved in the MPCD simulations.
Note, for large $\omega$ (data not shown) we observe deviations since the MPCD fluid becomes compressible\ \cite{Padding2006}.

\section{Taylor line in a microchannel}  \label{sec.micro}

In the following we present our simulation data of the Taylor line swimming in 
a microchannel and discuss it in detail.

\subsection{Swimming on a stable trajectory and under an acute angle at the channel wall}
\label{sec_micro}

\begin{figure}
 \includegraphics[width=0.42\textwidth]{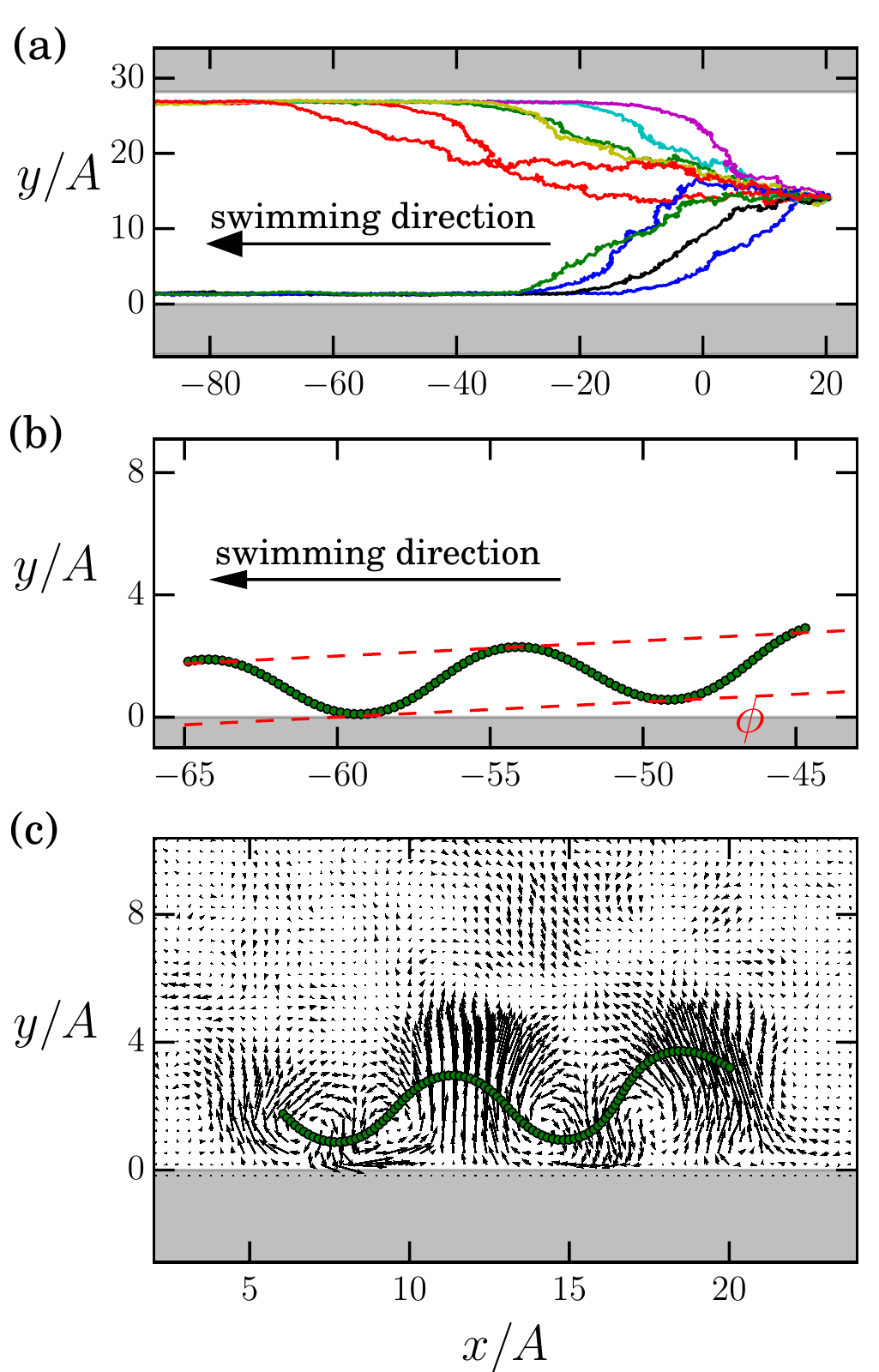}
\caption{Taylor lines swim along the walls of a microchannel (gray areas).
(a) Ten trajectories of the center of mass start in the middle and reach one of the walls.
Parameters are the channel width $d/A = 27.7$, the wave amplitude $A/\lambda = 0.1$, and
the wavelength $\lambda = 22.59 a_0$.
(b) Close-up: The Taylor line swims under an acute tilt angle $\phi$ along a channel wall.
(c) Close-up: Flow field initiated by the Taylor line when swimming along the channel wall 
}
\label{fig_micro_channel}
\end{figure}

In Fig.~\ref{fig_micro_channel} (a) we show ten center-of-mass trajectories of identical Taylor lines in a
wide microchannel with width $d/A = 27.7$.
They all start in the middle of the channel and always
swim in the negative $x$ direction towards one of the channel walls. After an axial swimming distance of
$80 A$,  $92\%$ of all our simulated Taylor lines have reached one of the channel walls (not all of the 
trajectories are shown here).
We observe that in
a very narrow channel with width $d/A = 3.07$, the swimming trajectory is not stable and the Taylor line 
switches from one wall to the other. However, already at $d/A = 3.75$ it  stays at one channel wall.
This occurs even though the walls are not further apart than 
four amplitudes. Stable swimming trajectories at channel walls have been observed in experiments and simulations 
of sperm cells and \text{E. coli} \cite{Rothschild1963, Elgeti2010, Berke2008}.

Figure\ \ref{fig_micro_channel} (b) shows that the  Taylor line swims 
at an acute tilt angle along the
channel wall. Earlier simulations of swimming sperm cells have attributed the attraction to the wall
to a pusher-like flow field, which drags fluid in at the sides of the swimmer \cite{Elgeti2010}. 
Thereby, the sperm cells are hydrodynamically attracted by the wall. Additional flow at the free end 
of the flagellum pushes the tail of the sperm cell up. In Fig.\ \ref{fig_micro_channel}(c) we 
confirm this picture.
Below the wave crests fluid is strongly pulled towards the Taylor line, while fluid flow 
towards the wall below the wave troughs 
is much weaker. Hence, the Taylor line is attracted to the wall. 
In addition, fluid flow towards the wave crest 
at the front is 
stronger compared to the second wave crest, which obviously 
tilts the Taylor line as Fig.\ \ref{fig_micro_channel} (b) demonstrates.
\begin{figure}
 \includegraphics[width=0.499\textwidth]{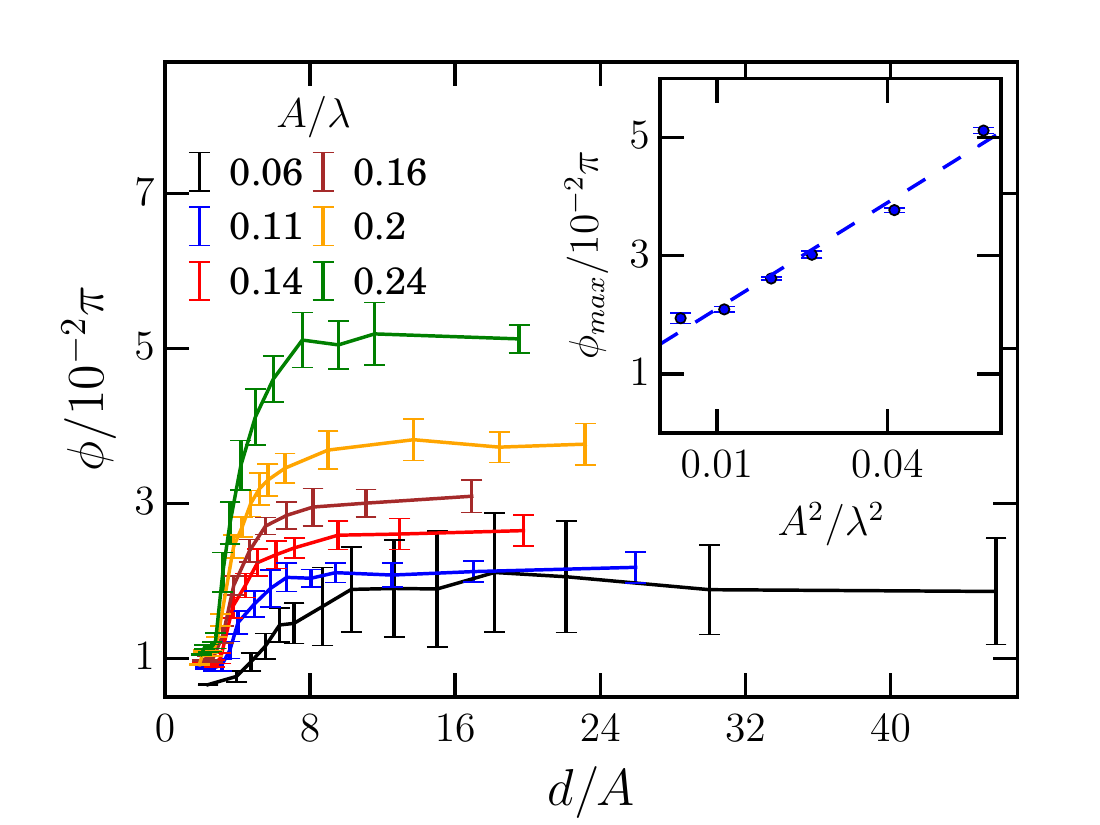}
\caption{
Mean tilt angle $\phi$ versus channel width $d/A$ for different amplitudes $A/\lambda$ 
at $\lambda = 21a_0$ and $\nu = 0.003 / \tau_0$.
Inset: Maximum tilt angle $\phi_{max}$ versus $(A /\lambda)^2$.
The dashed blue line is a linear fit to the data points.
}
\label{fig_angle} 
\end{figure}

In order to investigate the tilt angle $\phi$ at the channel walls in more detail, in Fig.~\ref{fig_angle}
we plot $\phi$ versus channel width for several amplitude-to-wavelength ratios $A/\lambda$. 
Each curve except for the smallest amplitude $A$ starts with a small region of the channel width $d/A \in [2,3]$, where
the tilt angle is ca. $0.01\pi$ and hardly depends on $d/A$.
Then, at the width $d/A \approx 3$ the tilt angle increases and
ultimately reaches a plateau value at  
$d/A \approx 8$ 
meaning that the Taylor line does not interact with the other channel wall
at widths $d/A \gtrsim 8$. 
The inset plots the plateau or maximum tilt angle $\phi_{max}$ versus $A^2/\lambda^2$.
It is determined as the average of all tilt angles for $d/A\gtrsim 8$.
The maximum tilt angle $\phi_{max}$ needs to be an even function in $A$ since $-A$ only introduces a phase shift of
$\pi$ in the bending wave, which does not change the steady state of the Taylor line. Indeed, we can fit our data by
\begin{align}
  \phi(A/\lambda) =  \phi_2 \frac{A^2}{\lambda^2}+ \phi_0 \, ,
\end{align}
where $\phi_2 = 1.944$ and $\phi_0 = 0.046$ are fit parameters.

\subsection{Speed enhancement at the channel wall}

\begin{figure}  
  \includegraphics[width=0.5\textwidth]{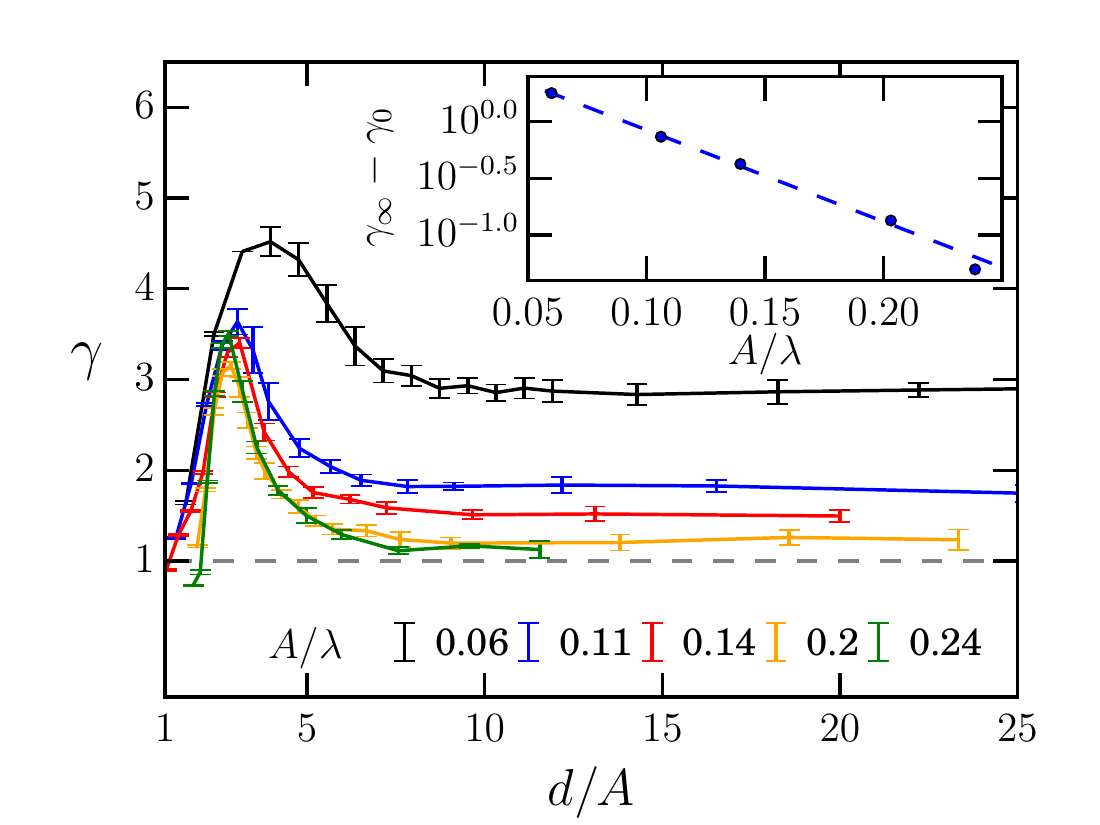}
 \caption{Speed enhancement versus dimensionless channel width $d/A$ for different amplitudes 
 $A/\lambda$. The inset plots log($\gamma_{\infty} - \gamma_0$) versus $A/\lambda$, where $\gamma_{\infty}$ is the plateau
 value and $\gamma_0$ a fit parameter.
 The dashed line shows an exponential fit to  $\gamma_{\infty} - \gamma_0 = \gamma_1 \exp(-\gamma_2 A/\lambda )$.
Fit parameters are $\gamma_0=1.08 \pm 0.03$, 
$\gamma_1= 5.4 \pm 0.3$, and $\gamma_2=-18.6 \pm 0.9$.
 }
 \label{fig_speed_channel_2} 
 \end{figure}

The swimming speed $\langle v_W \rangle$ of the Taylor line along the channel wall is enhanced compared
to the bulk value $\langle v_{\|} \rangle$  and strongly depends on the channel width. To 
discuss
this effect
thoroughly, we define a speed enhancement factor
\begin{equation}
 \gamma = \langle v_W \rangle  /  \langle v_{\|} \rangle \, ,
\end{equation} 
In Fig.\ \ref{fig_speed_channel_2}  
we plot 
it versus the channel width $d/A$.
Starting from $d/A \in [1,2]$, where the Taylor line squeezes into the channel,
$\gamma$ increases and goes through a maximum at $d/A \approx 3$. Interestingly, the maximum value of $\gamma$
is approximately the same, only for the smallest amplitude the maximum is larger and shifted towards $d/A \approx 4$. 
As before, at $d/A \gtrsim 8$ the factor $\gamma$ reaches a plateau value $\gamma_{\infty}$. 
Obviously, this happens when the other channel wall does no longer influence the swimming Taylor line by hydrodynamic
interactions. So the presence of both channel walls helps to speed up the Taylor line with an optimal channel width
at $d/A \approx 3$.

The inset shows how $\gamma_{\infty}$ decreases with increasing wave amplitude  $A$ and reaches nearly
one at $A/\lambda = 0.24$. This suggest the following interpretation. The Taylor line uses the no-slip condition of the fluid
at the channel wall to push itself forward. This is more effective the closer the Taylor line swims at the wall, \emph{i.e.},
for small $A$. In contrast, with increasing  $A$ also the mean distance of the Taylor line from the wall increases 
and one expects to reach the bulk value of the swimming speed ($\gamma_{\infty} = 1$) at large $A$.
The dashed line in the inset is an exponential
fit to $\gamma_{\infty} - \gamma_0 = \gamma_1 \exp(-\gamma_2 A/\lambda )$.
We find that $\gamma_0 = 1.08$  deviates
from the ideal large-amplitude value of one. This is due to a numerical artifact since for large $A$ the MPCD fluid is no longer
incompressible \cite{Padding2006}.

\section{Taylor line in a cubic obstacle lattice}
\label{sec_obstacel_lattice}

We now study the Taylor line swimming in a cubic lattice of obstacles with lattice constant $d$. 
Fig.\ \ref{fig_density_normal_swimming} shows the cubic unit cell. The obstacles have a diameter $2R / \lambda$, 
which we always refer to the wavelength $\lambda = 21a_0$ of the Taylor line. By varying $d$ and $R$, the 
Taylor line enters different swimming regimes, which we will discuss in detail in   
what follows.

\subsection{Dilute obstacle lattice}

\begin{figure}
 \includegraphics[width=0.5\textwidth]{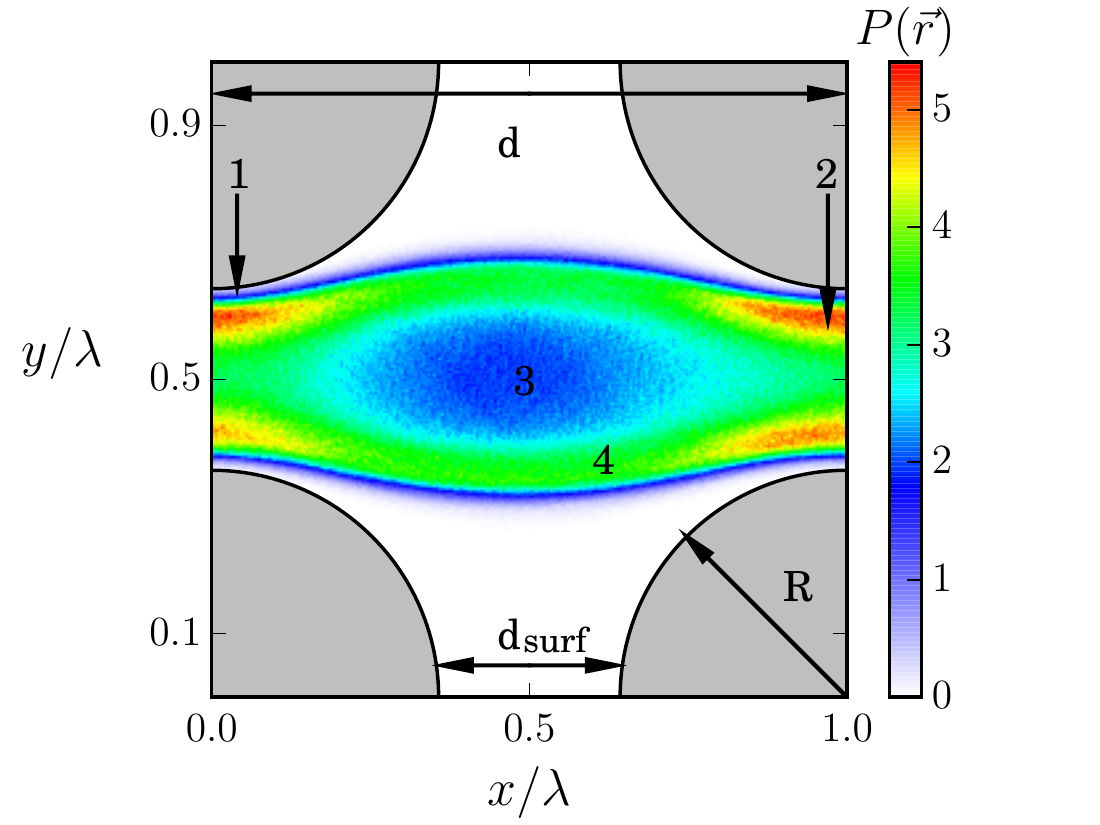}
\caption{Taylor line swimming in a dilute lattice of obstacles (gray quadrants).
The color code shows the probability density $P(\vec{r})$ for all bead positions of the Taylor line in the cubic unit cell
with lattice constant $d / \lambda = 1$, obstacle diameter $2R/\lambda = 0.714$, and gap width $d_{\textrm{surf}} = 2.04 A$.
The regions (1) - (4) are discussed in the main text.
}
\label{fig_density_normal_swimming}
\end{figure}

To define the dilute obstacle lattice, we introduce the width of the gap between two
neighboring obstacles,
\begin{align}
     d_{\textrm{surf}} = d - 2R \, .
\end{align}
For $d_{\textrm{surf}} > 2 A$
the Taylor line with amplitude $A$ can freely swim through the gap, whereas for $d_{\textrm{surf}} < 2 A$
it has to squeeze through the gap and therefore adjusts its swimming direction. This  leads to what we call 
geometrical swimming, which we will discuss in the following section.

We illustrate the first case, $d_{\textrm{surf}} > 2 A$, in Fig.\ \ref{fig_density_normal_swimming}, which shows 
the probability density $P(\vec{r})$ for all the beads of the Taylor line to visit a position $\vec{r}$ in the cubic unit cell. 
The probability density with the blue thin stripes shows that the Taylor line never leaves its lane. This is also true for 
other values of $d/\lambda$ as long as the Taylor line cannot freely rotate in the space between the lattices.
A closer inspection also shows a thin white region (1) around the obstacles, which the Taylor line never enters.  
Nevertheless, the probability of the beads for being in region (2) in the narrow gap between the obstacles is much 
higher than for being in region (3) between the four obstacles. 
We understand this as follows.
The beads move up and down while moving with the Taylor line.
In region (2) the beads reach their largest 
displacement equal to $A$ and slow down to invert their velocity. So, they spend more time in region (2), which explains 
the high residence probability not only in (2) but also in region (4). 

\begin{figure}
\includegraphics[width=0.49\textwidth]{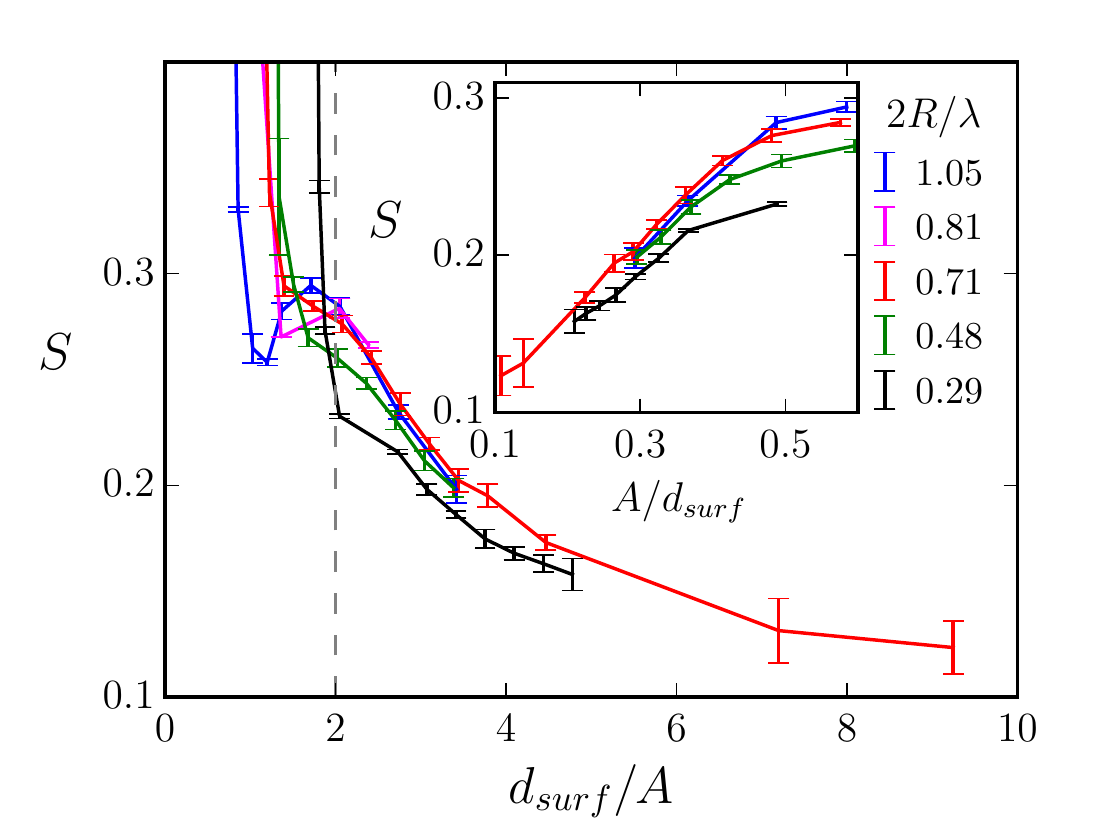}
\caption{Stroke efficiency $S$ plotted versus gap width $d_\textrm{surf}$ for different diameters
of the obstacles with $\lambda=21a_0$ and $A/\lambda = 0.14$. The vertical dashed line
separates the region of dilute ($d_{\textrm{surf}} > 2 A$) and dense ($d_{\textrm{surf}} < 2 A$) obstacle
lattices.
}
\label{d_surf_stroke} 
\end{figure}

In Fig.\ \ref{d_surf_stroke}
we plot the stroke efficiency as a function of $d_{\textrm{surf}}/A$ 
for different $2R/\lambda$. For $d_{\textrm{surf}}/A > 2$ the stroke efficiency ultimately is proportional 
to $1/d_{\textrm{surf}}$ as the inset demonstrates.
In addition, at constant $d_{\textrm{surf}}$ the efficiency $S$ is roughly the same, stronger deviations
only occur at the smallest $2R/\lambda = 0.29$. This means $S$ is mainly determined by the gap width,
through which the Taylor line has to move when $A$ is kept constant. 
For $d_{\textrm{surf}}<2A$ the Taylor line has to squeeze through
the obstacle lattice. 
In the main plot of Fig.\ \ref{d_surf_stroke} one
realizes a transition in all the curves,
where $S$ increases sharply. 
As we discuss in Sec.\ \ref{subsec_dense},
this is where the swimming Taylor line fits perfectly along one of the lattice directions and geometric swimming
takes place.

\subsection{Geometric swimming in a dense obstacle lattice}
\label{subsec_dense}

\begin{figure*}	
 \includegraphics[width=0.85\textwidth]{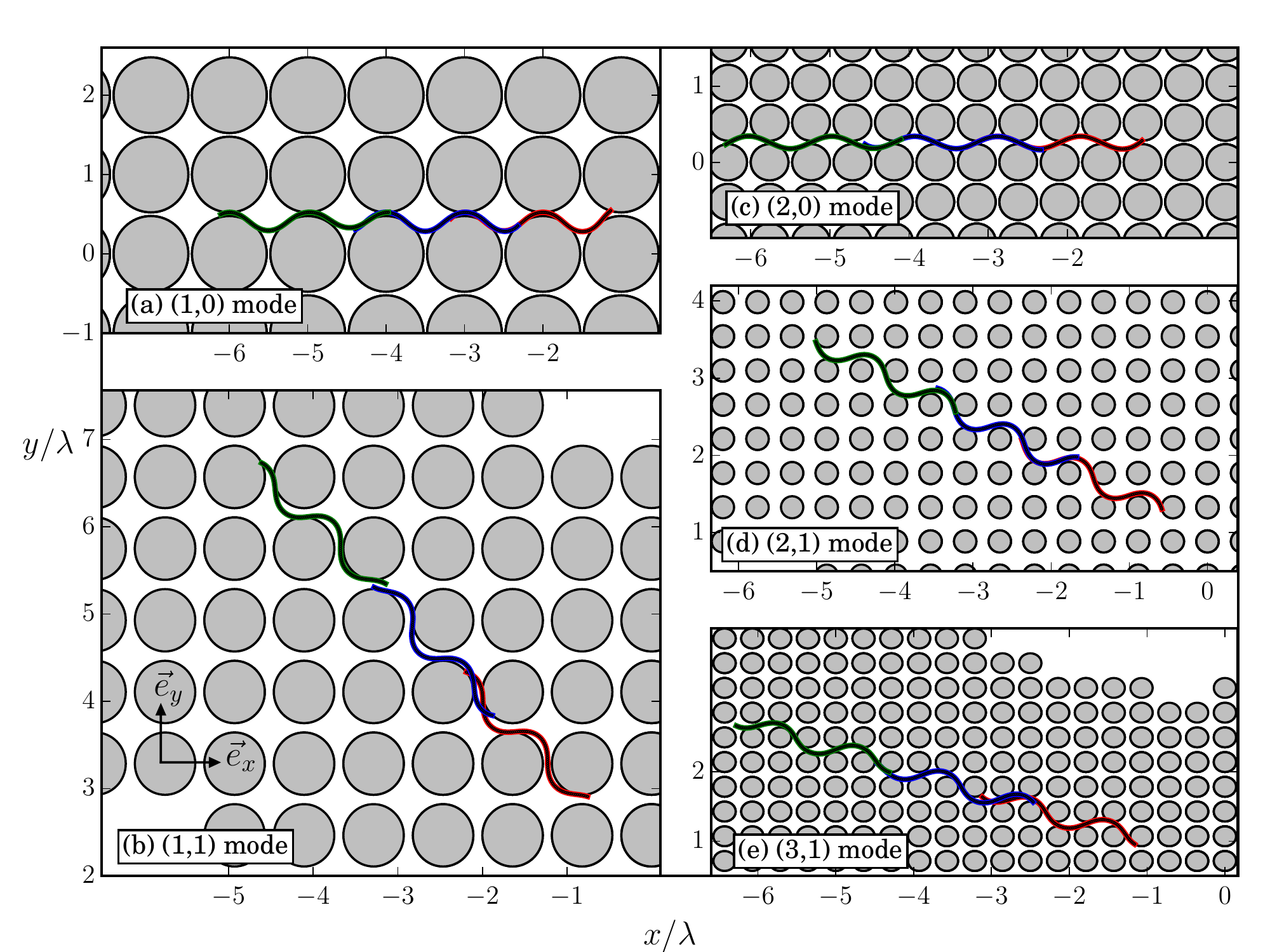}
\caption{Geometrical swimming of the Taylor line in a dense cubic lattice of obstacles (gray circles). Depending
on the lattice constant $d$, the Taylor line swims in different lattice directions with mode index $(m,n)$, where
$d(m \vec{e}_x + n \vec{e}_y)$ gives the direction of one wave train of the Taylor line and 
$\lambda \approx d \sqrt{m^2+n^2}$. 
Three snapshots with a 
time difference between $T$ and $2T$ are shown. The parameters of the illustrated swimming modes are:
(a) (1,0) mode with $d/\lambda = 0.95$ and $2R/\lambda = 0.95$, 
(b) (1,1) mode with $d_{\textrm{diag}} / \lambda = 1.08$ and $2R/\lambda = 0.71$, 
(c) (2,0) mode with $d/\lambda = 0.52$ and $2R/\lambda = 0.48$,
(d) (2,1) mode with $d/\lambda = 0.44$ and $2R/\lambda = 0.29$ 
[note $(2^2+1^2)^{-0.5} \approx 0.45$],
(e) (3,1) mode with $d/\lambda = 0.35$ and $2R/\lambda = 0.29$
[note $(3^2+1^2)^{-0.5} \approx 0.31$].
}
\label{trajek_diag}
\end{figure*}

In dense obstacle lattices ($d_{\textrm{surf}} < 2 A$) a new swimming regime occurs when the lattice
constant $d$ is appropriately tuned.
Starting 
to swim
in horizontal direction (see movie M1 in the supplemental material),
the Taylor line adjusts its swimming direction along a lattice direction with lattice vector $\vec{g} = d(m \vec{e}_x + n \vec{e}_y)$,
which defines the swimming mode $(m,n)$.
We call this regime geometrical swimming.
Figure\ \ref{trajek_diag} shows a few examples each with three snaphots of the Taylor line in green, red, and blue, where 
the time difference between the snapshots is between $T$ and $2T$. Perfect geometrical swimming occurs when one 
wave train fits perfectly into the lattice 
meaning 
\begin{equation}
\lambda = d_{\mathrm{eff}} = d \sqrt{m^2+n^2} \, ,
\label{eq.deff}
\end{equation}
where we have introduced the magnitude of the relevant lattice vector $d_{\mathrm{eff}} = |\vec{g}|$.
The (2,1) mode in the movie M1 
is a  good example
for geometric swimming. 
Depending on radius $R$ and amplitude $A$, the Taylor line 
also pushes against the obstacles. 
Obviously, for perfect geo\-metrical swimming the swimming velocity $v_{\parallel}$ and the phase velocity $c$ have to be identical: $v_{\parallel} = c$.
 The Taylor line swims with an efficiency $S=1$. It behaves like a corkscrew, which is twisted into a cork; 
after a full rotation the corkscrew has advanced by exactly one pitch. Differently speaking, the Taylor line converts the
bending wave optimally into a net motion without any slip between Taylor line and viscous fluid.
However, geometrical swimming also occurs when the perfect swimming condition is only approximately fullfilled, 
$\lambda \approx d \sqrt{m^2+n^2}$. In this case, the
Taylor line pushes against the obstacles and the swimming velocity deviates from $c$ 
but  
can even achieve values larger than $c$. 
We discuss this in the following.
Note that several of these swimming modes, in particular the (1,1) mode, have been observed in experiments for 
\textit{C. elegans} in an obstacle lattice 
\cite{Park2008, Majmudar2012}.

\begin{figure}
\includegraphics[width=0.5\textwidth]{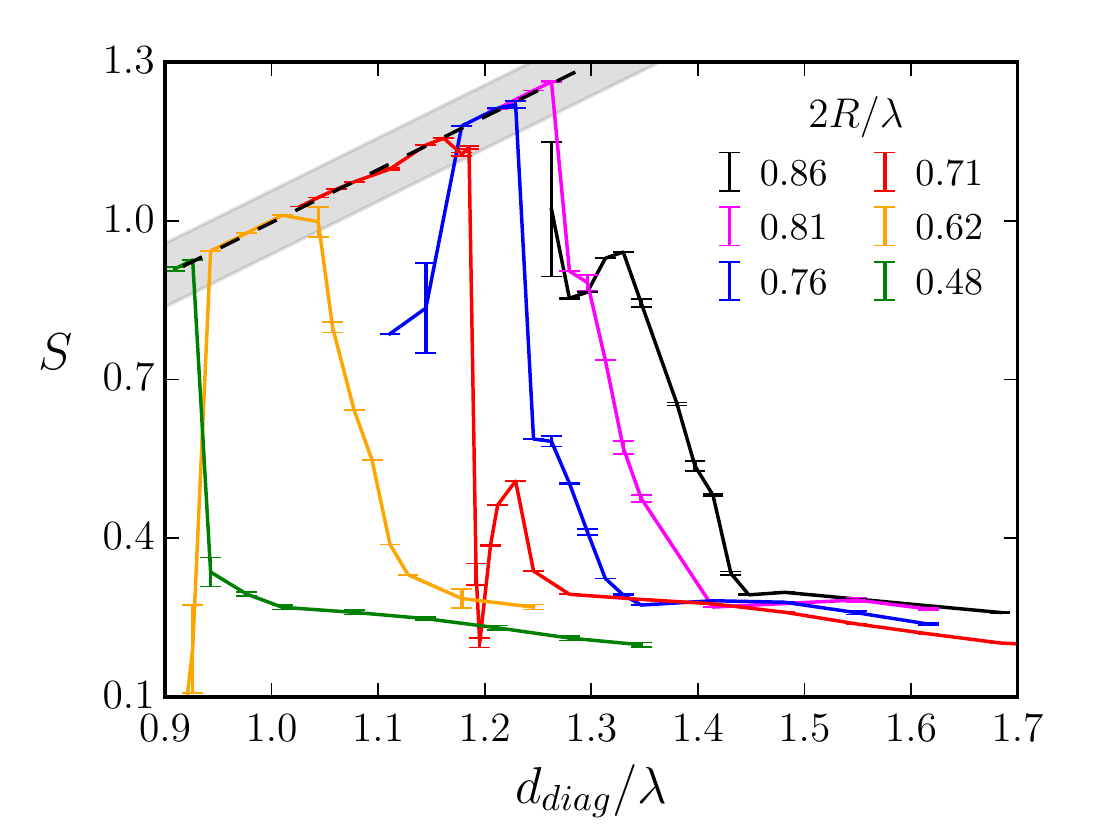} 
\caption{The stroke efficiency $S = v_{\|} / c$ for a Taylor line swimming predominantly in diagonal direction, i.e., in the
(1,1) mode. $S$ is plotted versus the diagonal distance $d_{\textrm{diag}}/\lambda$ between two obstacles for different 
obstacle diameters $2R/\lambda$. The gray shaded area shows the geometrical swimming regime and the dashed line 
with slope one indicates the geometric-swimming relation $S=d_{\textrm{diag}}/\lambda$ from Eq.\ (\ref{eq_stroke_d_eff}).
}
\label{fig_stroke_efficiency_dia}
\end{figure}

In the geometric swimming regime, the swimming efficiency $S = v_{\|} /c$ can be rewritten in pure geometric
quantities. Using $v_{\|} =  d_{\mathrm{eff}} \nu$ and $ c = \lambda \nu$, we immediately arrive at 
\begin{equation}
S = \frac{v_{\|}}{c} = \frac{d_{\textrm{eff}}}{\lambda} \, .
\label{eq_stroke_d_eff}
\end{equation}
In Fig.\ \ref{fig_stroke_efficiency_dia} we plot this relation as dashed line together with the gray shaded region to indicate
the geometric-swimming regime. The figure plots the stroke efficiency of a Taylor line swimming predominantly along
the diagonal direction in the lattice as a function of $d_{\mathrm{diag}}$, which is the diagonal distance of the obstacles. 
The curve parameter is the obstacle radius
$R/\lambda$. The sharp increase of $S$ in the orange curve ($2R/\lambda=0.62$) at $d_{\mathrm{diag}}=0.9$ indicates 
a transition from a swimming mode, where the Taylor line has to squeeze through the obstacle lattice,
to the geometric-swimming regime.
Then, a sharp decrease in $S$ follows
and ultimately $S$ decreases slowly. Increasing $d_{\mathrm{diag}}$ at constant $R$ makes the gaps between the 
obstacles wider and at the sharp decrease the Taylor line enters the regime of dilute obstacle lattices discussed in the 
previous section. 

\begin{figure}
 \includegraphics[width=0.5\textwidth]{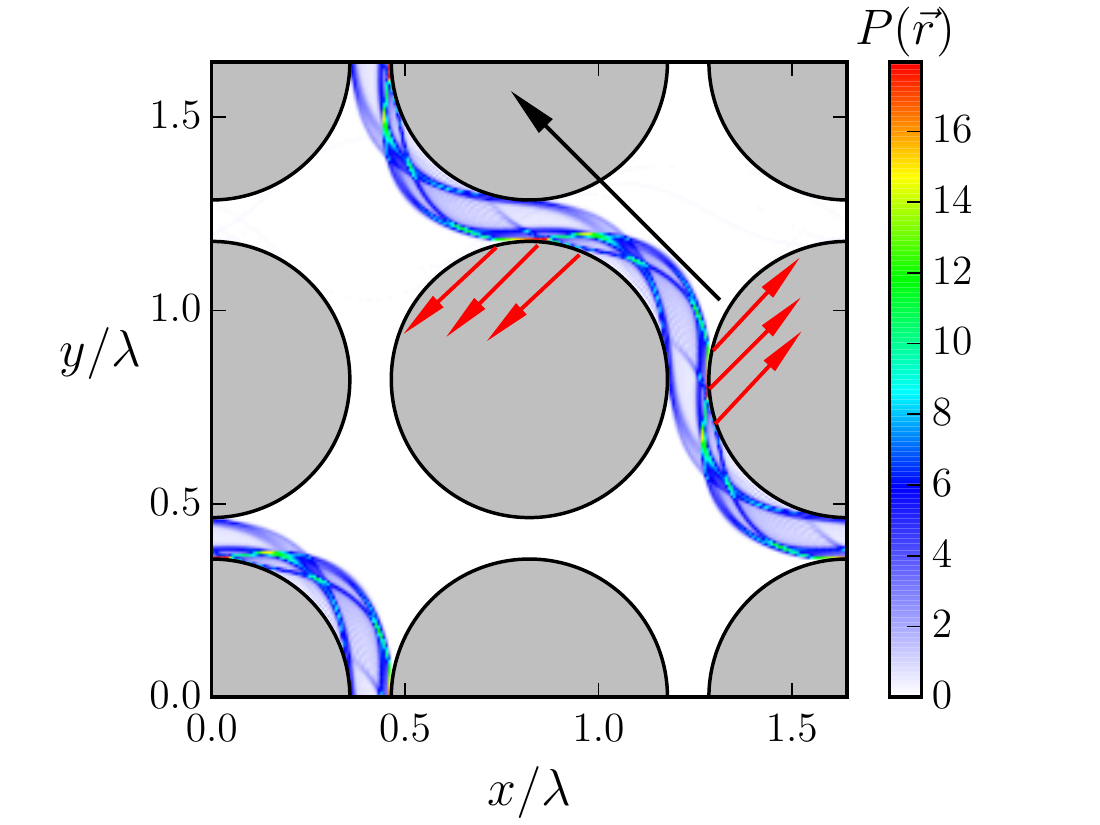}
\caption{Probability density $P(\vec{r})$ for all beads to visit a position in four unit cells during
geometrical swimming. The parameters are $d_{\textrm{diag}}/\lambda = 1.16$ and $2R/\lambda = 0.714$. 
The black arrow shows the swimming direction and the red arrows indicate where the head and middle section of 
the Taylor line push against the obstacles.
}
\label{fig_density} 
\end{figure}

The regime of geometric swimming extends over a finite interval in $d_{\mathrm{diag}}$.
One recognizes that geometric swimming can also be implemented when $d_{\mathrm{diag}} = \lambda$ is not exactly
fulfilled. Even swimming velocities larger than the wave velocity $c$ ($S>1$) are realized. Figure\ \ref{fig_density} illustrates the
mechanism for $d_{\mathrm{diag}} > \lambda$. It shows the probability density $P(\vec{r})$ summed over all beads to occupy a 
position between the obstacles. $P(\vec{r})$ reveals two sliding tracks of the Taylor line. A closer inspection shows that the 
head ($nl_0 \in [0,0.2L_c]$) and middle ($nl_0 \in [0.2L_c,0.7L_c]$) sections move on the ``pushing'' track. When the 
bending wave passes along the Taylor line, the Taylor line pushes against the obstacles (indicated by the red arrows), which helps it to swim faster than in the ideal case. This is nicely illustrated in movie M1 in the supplemental material for the (1,1) mode.
The other track is mainly occupied by the tail section ($nl_0 \in [0.7L_c,L_c]$) which does
not contribute to the increased propulsion. In between the tracks there is a blurry area indicating that the part of the Taylor line
between the middle and tail section has to transit from the pushing to the other track.

\begin{figure}
 \includegraphics[width=0.5\textwidth]{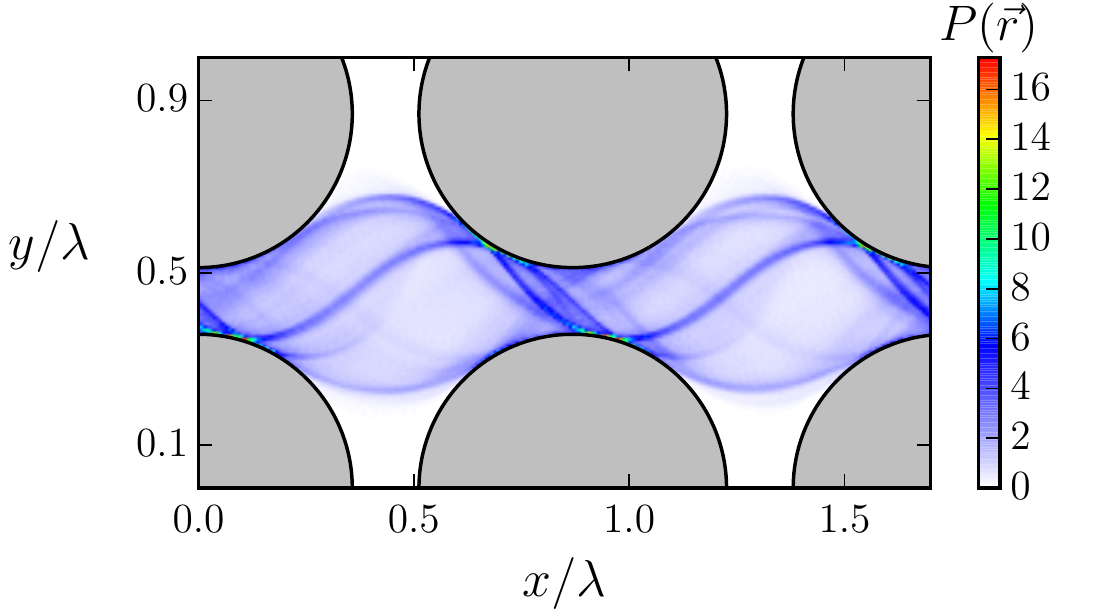}
\caption{Probability density $P(\vec{r})$ for all beads of the Taylor line to visit a position between the obstacles.
The Taylor line pushes against the obstacles. The parameters are $d_{\textrm{diag}}/\lambda = 1.23$ or 
$d/\lambda = 0,87$, $2R/\lambda = 0.714$, and $d_{\textrm{surf}}/A = 1.11$. 
}
\label{fig_density_second_peak}
\end{figure}

At larger obstacle diameters in Fig.\ \ref{fig_stroke_efficiency_dia} (red, blue, and purple line) the sharp decrease in $S$ 
after the geometric swimming indicates a different transition. The Taylor line changes direction and swims along the
(1,0) direction since then the wavelength $\lambda$ fits better to the spatial period, $\lambda \approx d$. The local maximum 
in the red curve develops into a shoulder, which for the purple curve belongs to the $(1,0)$ mode of geometrical swimming. 
Finally, for the black line ($2R/\lambda = 0.86$) geometric swimming along the (1,0) direction is more developed.
In Fig.~\ref{fig_density_second_peak} we show the positional probability density of all beads of the Taylor line exactly
at the local maximum of the red curve in Fig.\ \ref{fig_stroke_efficiency_dia}. With $d/\lambda = 0.87$ the Taylor line is not in
the geometric swimming regime. Even though the distribution is much more blurred than before, there is still a clear 
sinusoidal track visible. The Taylor line pushes against the obstacles, which helps it to move through the narrow
gap. Finally, the red curve in Fig.\ \ref{fig_stroke_efficiency_dia} becomes flat when the Taylor line enters the
dilute-lattice regime.

\begin{figure}
 \includegraphics[width=0.499\textwidth]{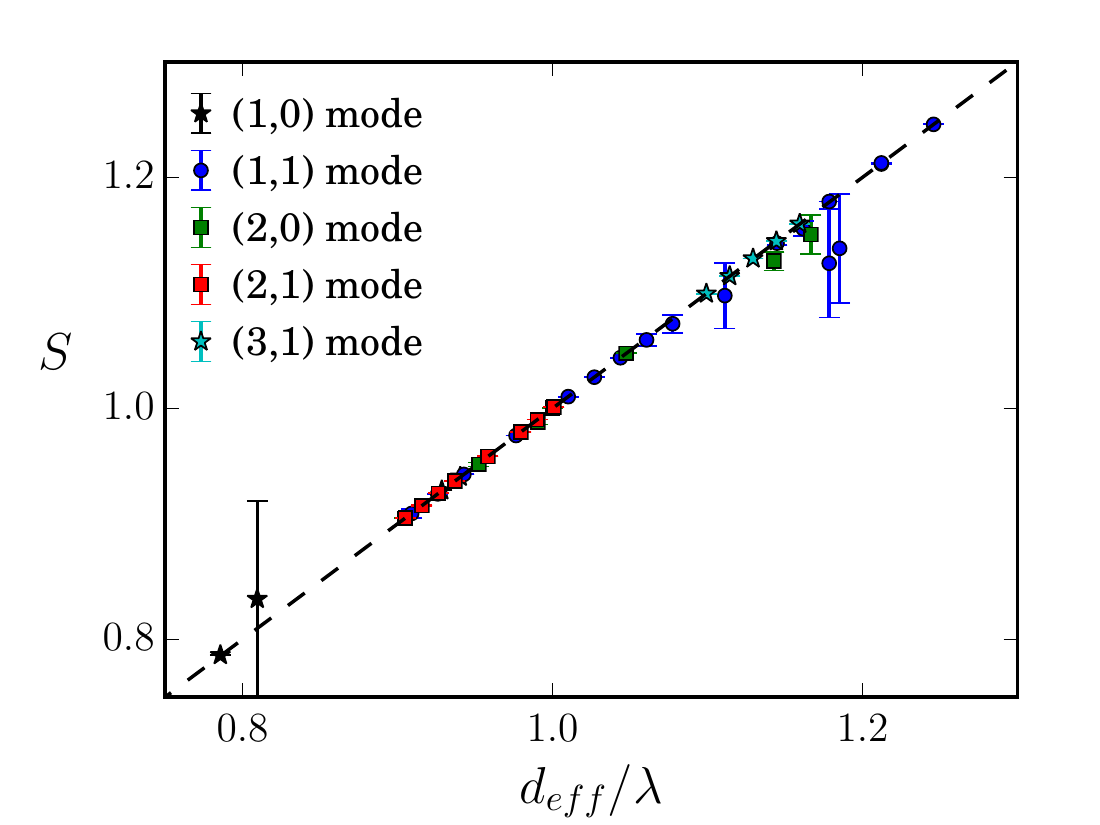}
\caption{Stroke efficiency $S$ versus effective distance $d_{\textrm{eff}}/\lambda$ defined in Eq.\ (\ref{eq.deff}) 
for different swimming modes $(m,n)$ and
for different parameters. All data in the geometrical swimming regime
collapse on one master curve.
}
\label{fig_master}
\end{figure}

For lattice constants $d$ well below $\lambda$ and smaller obstacle diameters $2R$, one also observes the higher modes
$(2,0)$, $(2,1)$, and $(3,1)$ visualized in Fig.\ \ref{trajek_diag}. In Fig.\ \ref{fig_master} we summarize all our
results by plotting $S$ for the different swimming modes against the specific $d_{\textrm{eff}}$ defined in Eq.\ (\ref{eq.deff}).
The resulting master curve impressively illustrates the significance of geometrical swimming even reaching swimming velocities
up to 20 \% larger than the ideal value from the phase velocity $c$. Thus, swimming in an obstacle lattice results in
a new type of swimming compared to conventional locomotion at small Reynolds numbers, it resembles rather a corkscrew
twisted into cork.

\subsection{More complex trajectories}

\begin{figure*}
 \includegraphics[width=0.85\textwidth]{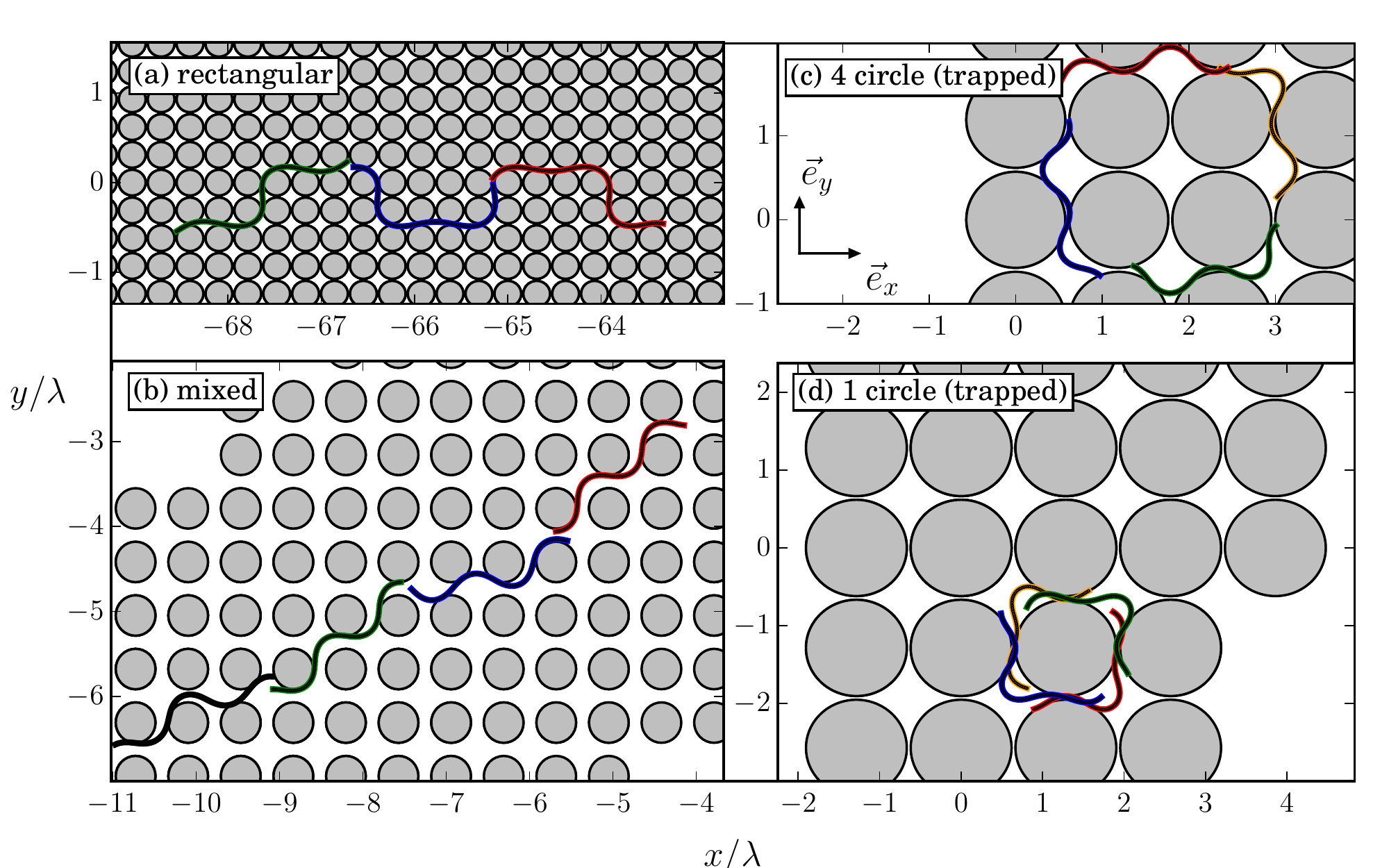}
\caption{
In a dense obstacle lattice more complex trajectories occur at specific values of lattice constant $d/\lambda$ and 
obstacle diameter $2R/\lambda$. Several snaphsots of the Taylor line are shown:
(a) rectangular mode at $d/\lambda = 0.31$ and $2R/\lambda = 0.29$; (b) mixed mode
at  $d/\lambda = 0.63$ and $2R/\lambda = 0.48$, where the Taylor line switches between the $(1,1)$ and $(3,1)$ 
swimming direction; (c) 4 circle (trapped) mode at $d/\lambda = 1.19 $ and $2R/\lambda = 1.14$, where the Taylor line circles
around four obstacles; and (d) 1 circle (trapped) mode at $d/\lambda = 1.29 $ and $2R/\lambda = 1.24$, where it circles
around one obstacle after an initial transient regime.
}
\label{fig_trajec_chaos}
\end{figure*}

In Fig.\ \ref{fig_trajec_chaos} we show examples of trajectories 
that do not show 
geometric swimming along a defined direction as discussed in Sec.\ \ref{subsec_dense}
but exhibit more complex shapes. They are also nicely illustrated in movie M2 of the supplemental material.
Depending on 
the specific values for lattice constant $d/\lambda$ and obstacle diameter $2R/\lambda$,
we can identify trajectories of different types. 
They either define new swimming modes [Fig.\ \ref{fig_trajec_chaos} (a), (c) and (d)] or combine two
geometric-swimming modes [Fig.\ \ref{fig_trajec_chaos} (b)].
In Fig.\ \ref{fig_trajec_chaos} (a) the obstacle lattice is so dense that the Taylor line cannot develop 
geometric swimming. Instead, it swims alternatively along the horizontal and vertical direction for four or two
lattice constants, respectively, which results in a trajectory of rectangular shape.
Figure\ \ref{fig_trajec_chaos} (b)
shows the Taylor line while it switches 
its running mode between the 
(1,1) and (3,1) swimming direction (see also movie M2).

A new trajectory type occurs when 
both the obstacle diameter $2R/\lambda$ and 
the lattice constant $d/\lambda$ 
roughly 
agree with the wavelength 
(see also movie M2).
In this case, after some transient regime the Taylor line is trapped and swims around a 
square of the same four obstacles [trapped circle mode in Fig.\ \ref{fig_trajec_chaos} c)] or around a single obstacle
[trapped circle mode in Fig.\ \ref{fig_trajec_chaos} d)].

\subsection{Variation of the length of the Taylor line}

\begin{figure}
 \includegraphics[width=0.5\textwidth]{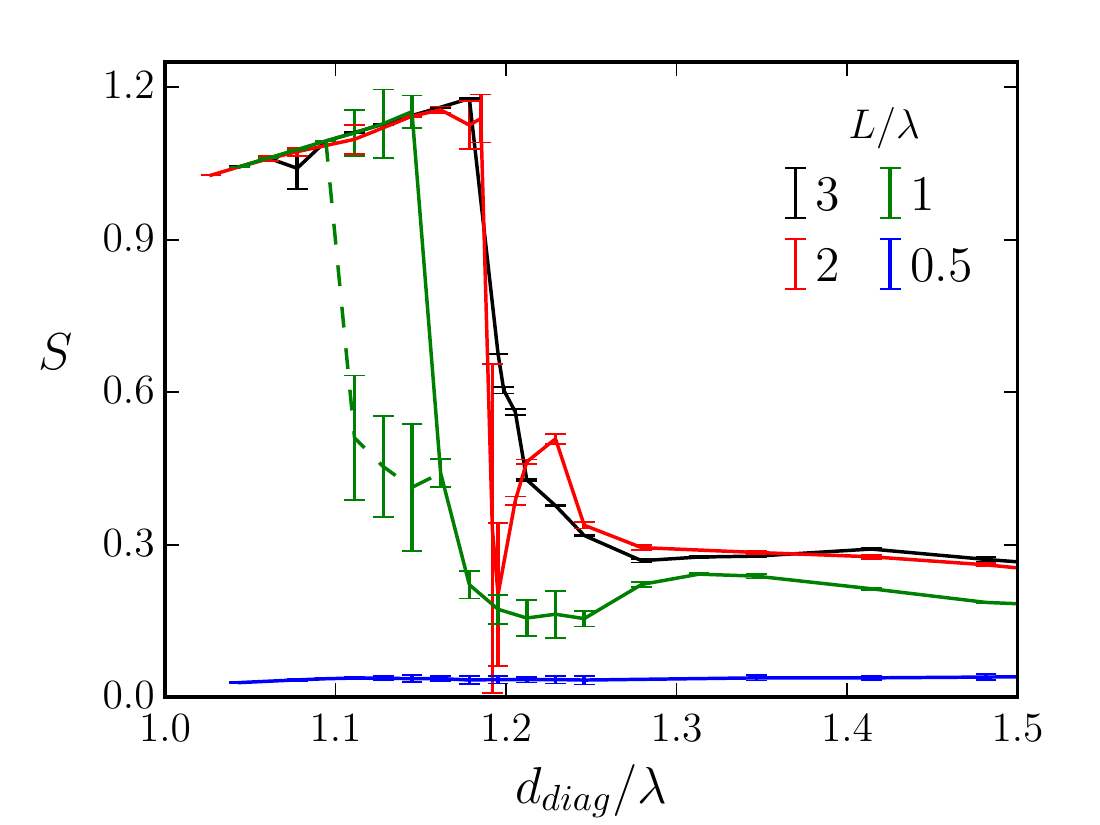}
\caption{Stroke efficiency $S$ versus diagonal distance $d_{\mathrm{diag}} / \lambda$ 
for different lengths $L / \lambda$ of the Taylor line at wavelength $\lambda = 21 a_0$ and obstacle
radius $R/\lambda=0.71$.
}
\label{speed_dif_length}
\end{figure}

In Fig.\ \ref{speed_dif_length} we plot the stroke efficiency $S$ versus diagonal obstacle distance $d_{\mathrm{diag}}$
for different lengths $L / \lambda$ of the Taylor line. We keep wavelength and obstacle radius constant. For 
$L / \lambda = 0.5$ the Taylor line hardly swims persistently, neither when it is strongly confined by the obstacles 
($d_{\mathrm{diag}} / \lambda < 1.3$) nor when it does not touch the obstacles at all ($d_{\mathrm{diag}} / \lambda > 1.3$).
This is nicely illustrated by movie M3. For $L / \lambda = 2$ and $3$ the Taylor lines first are clearly in the 
geometric-swimming regime along the $(1,1)$ direction. The strong decrease of $S$ at around $d_{\mathrm{diag}} / \lambda = 1.2$ 
indicates the transition to swimming along the $(1,0)$ direction. Right at the deep mimimum of the red curve ($L / \lambda = 2$)
the Taylor line gets more or less stuck before it enters the $(1,0)$ swimming direction. 
At ca. $d_{\mathrm{diag}} / \lambda > 1.4 $ the obstacles are sufficiently apart  from each other and the Taylor line does not push against them anymore.

At  length $ L / \lambda = 1$ and ca. $d_{\mathrm{diag}} / \lambda = 1.1$ a new feature occurs. 
The Taylor line switches between geometric 
swimming along $(1,1)$ and $(1,0)$ direction. This is illustrated by the two branches of the green curve in 
Fig.\ \ref{speed_dif_length} and in movie M4 for $d_{\mathrm{diag}} / \lambda = 1.13$. In the following broad
mimimum of the green curve ($1.17< d_{\mathrm{diag}}/\lambda < 1.25$), the Taylor line exhibits some stick-slip motion.
It first pushes frequently against
one obstacle and then swims more or less continuously for one lattice constant 
(see movie M4 for $d_{\mathrm{diag}} / \lambda = 1.2$). Again, at $d_{\mathrm{diag}} / \lambda > 1.4 $ the Taylor line
does not push anymore against the obstacles while swimming.

\section{Summary and Conclusions}
\label{sub_discus}

We have implemented an undulatory Taylor line in a Newtonian fluid using the method of multi-particle collision dynamics and
a sinusoidal bending wave running along the Taylor line. We have calibrated the parameters such that its peristence length is much 
larger than the contour length in order to observe regular undulatory shape changes and directed swimming.

In microchannels the Taylor line swims to one channel wall. Swimming speed is enhanced due to hydrodynamic
interactions and the Taylor line is oriented with an acute tilt angle at the wall similar to simulations of sperm cells \cite{Elgeti2010}.
The acute angle can be understood by monitoring the initated flow fields. 
In wide channels the tilt angle increases quadratically with the amplitude $A$ of the bending wave, while the speed enhancement decreases 
exponentially with increasing $A$ since the Taylor line swims, on average, further away from the wall. In narrow channels the 
swimming speed has a maximum at rougly $d/A \approx 3 $. The Taylor line uses the no-slip condition of the fluid at the walls
to effectively push itself forward.

In a dilute obstacle lattice swimming speed is also enhanced due to hydrodynamic interactions with the obstacles.
In the dense obstacle lattice we could reproduce the geometrical swimming observed in the case of \textit{C. elegans}
\cite{Park2008, Majmudar2012} even though we did not consider any finite
extension of
the Taylor line. 
In addition,
we found 
more complex swimming modes, which occur due to the strong confinement between the obstacles.
In 
the geometrical
swimming regime the Taylor line 
strongly interacts with the obstacles and swims with a speed close to the phase velocity of the bending wave,
thus much more efficiently than in a pure bulk fluid.
Geometrical swimming occurs when the wavelength of the Taylor line fits into the lattice along one specific direction.
Thus, the swimming efficiencies of various geometrical swimming modes, plotted versus the ratio $d_{\mathrm{eff}}/\lambda$
of effective obstacle distance and undulation wavelength, all collapse on the same master curve. Increasing
$d_{\mathrm{eff}}/\lambda$ beyond one, even swimming speeds larger than the phase velocity of the bending wave occur
but ultimately the Taylor line enters a different swimming mode. Thus, one can control the swimming direction of 
undulatory microorganisms by tuning the lattice constant of an obstacle lattice. This might be used for a microfluidic 
sorting device.

The concept of geometrical swimming goes back to Berg and Turner in order to explain the enhanced swimming of
helical bacteria in polymer networks of viscoelastic fluids \cite{berg1979}. Further studies on the undulatoryTaylor line should 
investigate the enhanced swimming speed in more disordered obstacle suspensions and when the obstacles are allowed to
move, which models more realistic environments such as blood.

\bibliographystyle{apsrev}
\bibliography{Diplomarbeit}


\begin{acknowledgments}
We acknowledge helpful discussions with C. Prohm, J. Blaschke, and A.  Z\"ottl. 
This research was funded by grants from DFG through the research training group GRK 1558
and project STA 352/9.
\end{acknowledgments}

\appendix

\section{Calibration of parameters}

\begin{table}

\begin{tabular}[c]{| r | l | r | r|} 
\hline\hline
$N$ & $b$ & $\lambda/a_0$ & $A/a_0$  \\
\hline
88 & 0.105  & 21.02 & 1.26 \\
\cline{2-4}
94 & 0.168 & 21.02 & 2.23 \\
\cline{2-4}
97 & 0.18725 & 20.99 & 2.60  \\
\cline{1-4}
100 & 0.06 & 24.40 & 0.94  \\
\cline{2-4}
100 & 0.15 & 22.59 & 2.27  \\
\cline{2-4}
100  & 0.2  & 20.99 & 2.93  \\
\cline{1-4}
105  & 0.2162   & 20.98 & 3.43  \\
\cline{2-4}
125  & 0.24 & 21.04& 5.02\\ \hline\hline
\end{tabular}

\caption{\label{tab_simu_param} Calibration of the parameters of the Taylor line.
The bead number $N$ and curvature parameter $b$ are the input parameters 
which determine the wavelength $\lambda$ and the amplitude $A$. Lengths are given in units
of the edge length $a_0$ of the collision cells.
}
\end{table}

We calibrate 
the amplitude $A$ and wavelength $\lambda$ of the Taylor line by varying the number of beads $N$ 
and the curvature parameter $b$. The parameters used in this article are summarized in Table\ \ref{tab_simu_param}.
The contour length is calculate by Eq. (\ref{eq_arclegth}). 
\end{document}